\documentclass{article}
\usepackage{arxiv}
\usepackage{fancyhdr}

\usepackage[utf8]{inputenc} 
\usepackage[T1]{fontenc}    
\usepackage{hyperref}       
\usepackage{url}            
\usepackage{booktabs}       
\usepackage{amsfonts}       
\usepackage{nicefrac}       
\usepackage{microtype}      
\usepackage{graphicx}
\usepackage{doi}
\usepackage[all]{xy}
\usepackage{setspace}
\usepackage[table]{xcolor}
\usepackage{comment}
\usepackage{amsmath}
\usepackage{xcolor}
\usepackage{multirow}
\usepackage{float} 

\usepackage{tikz}
\usetikzlibrary{cd}

\usepackage[shortlabels]{enumitem}

\title{Nonlinear GENERIC-Embedded Neural Networks (N-GENNs): learning GENERIC dynamics with non-quadratic dissipation potentials}

\author{Vojtěch Votruba\thanks{These authors contributed equally to this work.}\\
  Mathematical Institute, Faculty of Mathematics and Physics, Charles University,\\ 
  Prague, 18675, Czech Republic
 \And Zequn He\footnotemark[1]\\
  Department of Mechanical Engineering and Applied Mechanics, University of Pennsylvania,\\
  Philadelphia, 19104, PA, United States of America \\
 \And Weilun Qiu\\
 Department of Mechanical Engineering and Applied Mechanics, University of Pennsylvania,\\
 Philadelphia, 19104, PA, United States of America \\
 \And Celia Reina\thanks{Corresponding authors: \texttt{creina@seas.upenn.edu} (C.R.), \texttt{pavelka@karlin.mff.cuni.cz} (M.P.).}\\
 Department of Mechanical Engineering and Applied Mechanics, University of Pennsylvania,\\
 Philadelphia, 19104, PA, United States of America \\
 \And Michal Pavelka\footnotemark[2]\\
  Mathematical Institute, Faculty of Mathematics and Physics, Charles University,\\ 
  Prague, 18675, Czech Republic
 }

\date{}

\renewcommand{\headerleft}{N-GENNs}
\renewcommand{\headerright}{\doi{10.1016/j.cma.2026.119277}}

\newcommand{\bb}{\mathbf{b}}
\newcommand{\aaa}{\mathbf{a}}
\newcommand{\DD}{\mathrm{D}}
\newcommand{\xx}{\mathbf{x}}
\newcommand{\nn}{\mathbf{n}}
\newcommand{\dd}{\mathrm{d}}
\newcommand{\xxstar}{\mathbf{x}^{*}}

\begin{document}
\maketitle

\begin{abstract}
We introduce Nonlinear GENERIC-Embedded Neural Networks (N-GENNs), a deep learning framework for discovering evolution equations of systems governed by the nonlinear GENERIC formalism (General Equation for Non-Equilibrium Reversible–Irreversible Coupling). Such systems exhibit coupled conservative and dissipative dynamics, and can be described via the superposition of a Hamiltonian flow and a generalized gradient flow. In contrast to existing approaches, our formulation incorporates generalized gradient flows via convex dissipation potentials, enabling the identification of a broader class of thermodynamically consistent dynamics, including systems with non-quadratic dissipation potentials. Thermodynamic structure is strongly enforced by construction through suitable reparameterizations of both the reversible operator and the dissipation potential, ensuring exact compliance with the first and second laws of thermodynamics. We validate the proposed approach on three representative examples: a harmonic oscillator coupled to a heat bath, an idealized chemical motor, and a one-dimensional viscoplastic model of Perzyna type. These results demonstrate the method’s ability to accurately infer thermodynamically consistent models from data for systems incorporating both conservative and nonlinear dissipative dynamics.
\end{abstract}


\tableofcontents

\section{Introduction}
\label{sec:intro}
Data-driven discovery of governing equations has emerged as a powerful paradigm for modeling complex physical systems whose dynamics are difficult to derive from first principles alone. Early approaches such as Sparse Identification of Nonlinear Dynamics (SINDy) \cite{brunton2016discovering} have demonstrated that interpretable evolution laws can be extracted directly from measurement data, while Neural Ordinary Differential Equations (Neural ODEs) \cite{chen2018neural} and Physics-Informed Neural Networks (PINNs) \cite{raissi2019physics} have shown that neural networks can serve as a flexible surrogates for both forward and inverse problems governed by differential equations. Despite these advances, unconstrained data-driven models may reproduce results that violate fundamental physical principles. Such inconsistencies are particularly consequential in computational mechanics, where reliable long-time predictions often require that the learned dynamics remain compatible with conservation laws, irreversible entropy production, and the geometric structure of the underlying physics. 

These challenges have motivated the development of structure-preserving machine learning, a paradigm that seeks to embed physical and mathematical structure directly into the learned models. In contrast to physics-informed approaches \cite{karniadakis2021physics}, which typically enforce physical laws weakly through penalty terms in the loss function, structure-preserving methods aim to guarantee these properties by construction (often via explicit constraints or carefully designed model architectures), although hybrid hard- and soft-constrained formulations have also been explored \cite{celledoni2021structure}\footnote{We remark that the terminology in the literature is not uniform. In this work, we reserve the term ``structure-preserving'' for approaches that enforce the underlying constraints exactly (hard constraints), and distinguish them from methods that impose such constraints only approximately through penalty terms.}.

A significant body of research in structure-preserving machine learning has been focused on reversible dynamics. Hamiltonian Neural Networks (HNNs) \cite{greydanus2019hamiltonian} learn the scalar Hamiltonian and recover the dynamics through Hamilton's equations, while Lagrangian Neural Networks (LNNs) exploit the variational formulation of the Euler-Lagrange equations \cite{cranmer2019lagrangian}. Symplectic architectures constrain the learned flow to preserve symplecticity and often exhibit improved long-time behavior \cite{jin2020sympnets, zhong2020symplectic}. For non-canonical conservative systems, Poisson Neural Networks (PNNs) \cite{jin2022learning}, Direct Poisson Neural Networks (DPNNs) \cite{vsipka2023direct}, and Lie--Poisson Neural Networks \cite{eldred2024lie} extend the same philosophy to systems with nontrivial Poisson brackets and Casimir invariants. 

Collectively, these approaches are primarily designed for isolated Hamiltonian systems, where the dynamics preserve energy and entropy. However, many systems of practical interest in mechanics, such as frictional and viscoplastic solids, complex fluids, and reactive media, exhibit an intrinsic coupling between reversible and dissipative processes, leading to irreversible entropy production \cite{gay2018lagrangian, giga2017variational, mielke2011formulation, zafferi2023generic}. Constructing physically consistent models for such systems therefore requires frameworks that incorporate both reversible and dissipative effects while remaining compatible with the first and second laws of thermodynamics.

\begin{sloppypar}
A modern framework for such coupled dynamics is the General Equation for Non-Equilibrium Reversible--Irreversible Coupling (GENERIC), introduced by Grmela and {\"O}ttinger \cite{grmela1997dynamics, ottinger1997dynamics}. The framework combines Hamiltonian and gradient flow structures into a single thermodynamically consistent formalism \cite{dzyaloshinskii1980poisson, grmela1984particle, morrison1984bracket, kaufman1984dissipative}, and has been further developed in the non-equilibrium thermodynamics literature, including two monographs \cite{pavelka2018multiscale, ottinger2005beyond}. In GENERIC, the evolution of a state vector $\xx$ is generated by the Poisson operator $L(\xx)$, the energy functional $E(\xx)$, the dissipation potential $\Xi(\xx, \xxstar)$, and the entropy functional $S(\xx)$ as

\begin{equation}\label{Eq:Nonlinear_GENERIC}
    \dot{\xx} = L(\xx)\DD E(\xx) + \DD_{\xxstar} \Xi(\xx,\xxstar)|_{\xxstar = \DD S(\xx)}.
\end{equation}
This formulation is often referred to as nonlinear GENERIC, as the dissipative contribution is generated by a generally non-quadratic dissipation potential $\Xi$, in which case $\DD_{\xxstar} \Xi$ is non-linear with respect to $\xxstar$.
The linear formulation of GENERIC, equally well recognized in the literature and tightly connected to the older idea of so-called metriplectic evolution \cite{morrison1984bracket}, replaces the dissipation potential by a friction matrix $M(\xx)$, as
\begin{equation*}
    \dot{\xx} = L(\xx)\DD E(\xx) + M(\xx) \DD S(\xx).
\end{equation*}
In both cases, the building blocks are further subjected to degeneracy conditions that enforce non-negative entropy production and energy conservation (this will be made precise in Section \ref{sec:generic-formalism}). 

Owing to its structure, GENERIC has become a versatile modeling framework across complex fluids, thermoelasticity, dissipative solids, and reactive systems \cite{mielke2011formulation,  zafferi2023generic}. Its built-in thermodynamic consistency also makes GENERIC a natural inductive bias for data-driven modeling. Consequently, a growing body of work in thermodynamics-informed machine learning seeks to combine GENERIC structure with neural network parameterizations. Both soft- and hard-constrained approaches have been explored in this context. Soft-constrained methods penalize violations of the degeneracy conditions of GENERIC rather than enforcing them exactly \cite{hernandez2021structure}. In contrast, hard-constrained approaches encode thermodynamic admissibility in the parameterization itself \cite{lee2021machine, zhang2022gfinns, hernandez2022thermodynamics, gruber2023reversible, gruber2025efficiently, hernandez2025data}.

\end{sloppypar}
Within neural network formulations, current approaches broadly follow two directions: learning the full coupled GENERIC dynamics using the dissipative operator $M$, or focusing on formulations that consider general dissipation potentials $\Xi$, including quadratic and non-quadratic models.

Along the first direction, several works have aimed to learn thermodynamically structured dynamics directly from trajectory data. GNODE \cite{lee2021machine} introduced a data-driven parameterization of structure-preserving dissipative brackets for forecasting irreversible processes, in which the operators $L(\xx)$ and $M(\xx)$ are modeled as functions of $\DD S(\xx)$ and $\DD E(\xx)$, respectively. A potentially more general framework, GFINNs \cite{zhang2022gfinns}, constructs the reversible and irreversible operators through structured reparameterizations that are sufficiently expressive to approximate arbitrary $L(\xx)$ and $M(\xx)$. 

Subsequent metriplectic formulations have extended these ideas to energetically consistent model reduction \cite{gruber2023energetically}, graph neural network architectures informed by reversible and irreversible bracket dynamics \cite{gruber2023reversible}, efficiently parameterized neural metriplectic systems with improved scalability \cite{gruber2025efficiently}, and structure-preserving coarse-graining for particle systems based on discrete metriplectic dynamics \cite{hernandez2025data}. Port-metriplectic neural networks \cite{hernandez2023port} further generalize this framework to open systems by enabling modular learning of subsystems interconnected through energy ports. A common issue, however, is that while the dynamics may be unique from purely (deterministic) macroscopic data, the thermodynamic structure is, in general, not unique \cite{huang2022variational, zhang2022gfinns}. This issue was resolved for purely dissipative systems by Stat-PINNs \cite{huang2025statistical}, which further exploit the information contained in the fluctuations of the macroscopic fields through a fluctuation-dissipation relation to identify the unique dissipative operator together with the dissipative force, yielding an interpretable coarse-graining strategy. 

As part of the second direction, Variational Onsager Neural Networks (VONNs) \cite{huang2022variational} learn isothermal dissipative dynamics in variational form based on Onsager's variational principle \cite{doi2011onsager,arroyo2017onsager}, by parameterizing the evolution through a learned free energy and dissipation potential. In particular, the dissipation potential is represented using a partially-input-convex neural network, so that non-negative dissipation and thus the second law of thermodynamics are enforced by construction (further details will be provided in Section \ref{sec:generic-formalism}). The resulting variational formulation can represent nonlinear, non-quadratic dissipative behavior, of relevance in generalized standard materials in solids, generalized diffusion phenomena and chemical reactions. EUCLID \cite{flaschel2023automated} demonstrates the value of this formulation in solid mechanics by automatically discovering generalized standard material models from a parametric library of material models by leveraging sparsity promoting regularization strategies. More recently, Generalized Standard Material Networks introduced in \cite{flaschel2025convex} have applied this approach to direct constitutive learning of viscoelastic, elastoplastic and
viscoplastic models with different hardening mechanisms. Furthermore, \cite{tacc2023data,holthusen2024theory} explore applications within the finite kinematics setting. 

Despite this progress, an important gap remains. To the best of our knowledge, there is currently no framework for discovering the full set of GENERIC building blocks in the dissipation-potential formulation, as in Eq.~\eqref{Eq:Nonlinear_GENERIC}. This is an important omission, as non-quadratic dissipation potentials are essential for modeling complex irreversible mechanisms exhibiting non-Gaussian fluctuations, such as chemical reactions, jump processes, or plasticity \cite{mielke2014relation, ottinger2021framework, kraaij2020fluctuation}. Although, from a deterministic perspective, formulations based on a dissipative matrix can be seen as more general than GENERIC with a dissipation potential \cite{hutter2013quasi, ottinger2019combined}, the latter is seen as exhibiting wider generality when considering fluctuations, as noted in the large deviation literature \cite{ottinger2021framework}. Parameterizing the dissipation potential $\Xi$ also offers important advantages from both modeling and learning perspectives. First, it leads to a more parsimonious representation: instead of learning a matrix-valued operator, one only needs to approximate a scalar function $\Xi(\xx, \xxstar)$. As a consequence, the thermodynamic constraints reduce to simple properties of $\Xi$ (non-negativity, convexity, and vanishing at the origin) rather than requiring the enforcement of positive semi-definiteness of a matrix field. Second, the dissipation potential admits a direct statistical-mechanical interpretation, as it is related to the rate function in the large deviation principle governing fluctuations around the most probable trajectory \cite{mielke2014relation}. This connection has already been exploited in purely dissipative systems to uniquely identify the dissipation potential from macroscopic fluctuation data \cite{ottinger2021framework,montefusco2021frameworkb}.

To address the aforementioned gap, we propose Nonlinear GENERIC-Embedded Neural Networks (N-GENNs), a machine learning framework for GENERIC systems with general dissipation potentials, applicable to both quadratic and non-quadratic cases, as in Eq.~\eqref{Eq:Nonlinear_GENERIC}. N-GENNs combine a learned reversible structure with a learned dissipation potential and entropy, and enforce the thermodynamic constraints by construction thereby ensuring exact compliance with the first and second laws of thermodynamics. We demonstrate the capabilities of N-GENNs on three representative examples: a harmonic oscillator coupled to a heat bath, an idealized chemical motor, and a one-dimensional Perzyna-type viscoplastic model. 

The main contributions of this work can thus be summarized as
\begin{itemize}
\item A deep learning framework for discovering evolution equations governed by the nonlinear GENERIC formalism, in which the energy, entropy, reversible operator, and dissipation potential are parameterized by neural networks whose architectures hard-encode the required structural properties to ensure satisfaction of the thermodynamic laws by construction.
\item A validation of the proposed framework on three benchmark systems spanning ODEs and PDEs, smooth and non-smooth dissipation potentials, and mechanical and multiphysics problems.
\end{itemize}

The remainder of the paper is organized as follows. Section~\ref{sec:generic-formalism} reviews the nonlinear GENERIC formalism. Section~\ref{sec: nn-generic} introduces the N-GENNs architecture. Section~\ref{sec:numerical-examples} presents the numerical examples for the three aforementioned examples, described by either ordinary or partial differential equations. Finally, the paper concludes with a discussion of the main findings, its limitations, and directions for future work.

\section{GENERIC formalism}
\label{sec:generic-formalism}
As introduced previously, the GENERIC formalism was first developed in \cite{ottinger1997dynamics, grmela1997dynamics}, building upon earlier efforts to couple Hamiltonian mechanics with dissipative dynamics \cite{dzyaloshinskii1980poisson, kaufman1984dissipative, morrison1984bracket, grmela1984particle, beris1994thermodynamics}. 
It provides a unifying and thermodynamically consistent framework for modeling a wide class of non-equilibrium dynamics, including mechanical systems with friction, chemical reactions, fluids, mixtures, elasticity, plasticity, non-Newtonian fluids, or kinetic theory \cite{ottinger1997dynamics, grmela1997dynamics, ottinger2005beyond, pavelka2018multiscale, mielke2011formulation}. In its classical formulation, the GENERIC framework applies to closed systems, for which total energy is conserved. It is nevertheless frequently extended to open systems by adopting an enlarged state‑space description, in which the state variables account for both the system and its surrounding reservoirs.

For an evolution of a general state vector~$\xx$, which may be finite or infinite dimensional (i.e., a field), GENERIC assumes that the dynamics is decomposed into a reversible Hamiltonian part and an irreversible generalized gradient flow,
\begin{equation}\label{eq:generic}
    \dot{\xx} = L(\xx)\DD E(\xx) + \DD_{\xxstar} \Xi(\xx,\xxstar)|_{\xxstar = \DD S(\xx)}.
\end{equation}
Here, the Hamiltonian $E(\xx)$, the entropy $S(\xx)$, and the dissipation potential $\Xi(\xx, \xxstar)$ can then either represent infinite-dimensional field functionals or finite-dimensional functions. Depending on the setting, the $\DD$ operator represents a functional derivative or a partial derivative. In more general situations where $\Xi$ is not differentiable, the convex subdifferential must be employed, and the resulting dynamics are formulated as a differential inclusion \cite{mielke2011formulation}. In the following, however, we restrict our attention to the differentiable case for the sake of simplicity, or suitable unique representations of $\DD_{\xxstar} \Xi$. The Poisson operator $L(\xx)$ is a 2 times contravariant tensor field encoding the reversible structure of the dynamics and may equivalently be interpreted in terms of a Poisson bracket,
 $\{V,W\}= (\DD V)^T L (\DD W)$.

Typically, six constraints are imposed on these functions from which the fulfillment of the thermodynamic laws follows (this will be shown next) \cite{ottinger1997dynamics, grmela2018generic, kraaij2020fluctuation, ottinger2005beyond, pavelka2018multiscale}:
\begin{enumerate}[(i)]
    \item The reversible degeneracy condition $L(\xx)\DD S(\xx) = 0, \quad \forall \xx$. \label{constr:1}
    \item The irreversible degeneracy condition $\Xi(\xx, \xxstar + \lambda \DD E (\xx)) = \Xi(\xx, \xxstar), \quad \forall \xx, \xxstar \quad \forall \lambda \in \mathbb{R}$. \label{constr:2}
    \item $\Xi(\xx, 0) = 0$ and $\Xi(\xx, \xxstar) \geq 0, \quad \forall \xx, \xxstar$. \label{constr:3} 
    \item Convexity of $\Xi(\xx, \xxstar)$ with respect to $\xxstar$ everywhere. \label{constr:4}
    \item The skew-symmetry of the Poisson operator $L(\xx) = -L^T(\xx), \quad \forall \xx \label{constr:5}$.
    \item The Jacobi identity for the Poisson operator: $\sum_k (L^{kl}\DD_k L^{ij} + L^{ki}\DD_k L^{jl} + L^{kj} \DD_k L^{li}) = 0, \forall\ l, i, j$, meaning that the Lie derivative of the Poisson operator with respect to the Hamiltonian vector field vanishes \cite{fecko2006differential}. The indices here denote the components of the Poisson operator and the derivatives with respect to specific state variables. \label{constr:6}
\end{enumerate}

An important class of dissipation potentials consists of those that are quadratic in their second argument, of the form $\Xi(\xx, \xxstar) = \frac{1}{2} (\xxstar)^T M(\xx) \xxstar$. Here, $M$ is the dissipative matrix, whose symmetry encodes the celebrated Onsager reciprocal relations \cite{onsager1931reciprocal,onsager1931reciprocalb}.

\paragraph{\label{sec:2} Conservation laws and convexity}
As an immediate consequence of \eqref{constr:1}, \eqref{constr:3} and \eqref{constr:4} the GENERIC evolution maximizes entropy. Indeed, by the chain rule, we obtain
\begin{equation}
    \frac{\dd S}{\dd t} = \DD S^T \dot{\xx} = \DD S^TL(\xx)\DD E(\xx) + \DD S^T\DD_{\xxstar} \Xi(\xx,\xxstar)|_{\xxstar = \DD S(\xx)} = (\DD S)^T \DD_{\xxstar} \Xi(\xx,\xxstar)|_{\xxstar = \DD S(\xx)} \geq 0. \label{eq:entropy_max}
\end{equation}
The last inequality is true for any convex function with minimum at the origin. In recent years, however, non-convex dissipation potentials that still lead to entropy maximization have been employed \cite{janevcka2018non}. Instead of global convexity, one then imposes only local convexity around $\xxstar = 0$ (obtaining equilibrium stability) and the so-called radial monotonicity which makes \eqref{eq:entropy_max} non-negative \cite{pavelka2018multiscale}.

The second consequence we aim to achieve is energy conservation. Using \eqref{constr:5},
\begin{equation}
    \frac{\dd E}{\dd t} = \DD E^T \dot{\xx} = \DD E^T L(\xx)\DD E(\xx) + \DD E^T\DD_{\xxstar} \Xi(\xx,\xxstar)|_{\xxstar = \DD S(\xx)} = (\DD E)^T \DD_{\xxstar} \Xi(\xx,\xxstar)|_{\xxstar = \DD S(\xx)}.
\end{equation}
Finally, the right-hand side of this expression vanishes, which is apparent from differentiating \eqref{constr:2} by $\lambda$, and plugging in $\lambda = 0$. Analogously, one can impose more constraints to enforce the conservation of other quantities, if applicable, such as linear momentum or the total mass of the system \cite{grmela2018generic}.

Nevertheless, from the expression above, it becomes apparent that one could think of at least three ways to satisfy the conservation of energy
\begin{enumerate}[(A)]
    \item \label{A} $\Xi(\xx, \xxstar + \lambda \DD E(\xx)) = \Xi(\xx, \xxstar), \quad \forall \xx,\xxstar, \quad \forall \lambda \in \mathbb{R}$ 
    \item \label{B} $\Xi(\xx, \DD S(\xx) + \lambda \DD E(\xx)) = \Xi(\xx, \DD S(\xx)), \quad \forall \xx, \quad \forall \lambda \in \mathbb{R}$ 
    \item \label{C} $\DD E^T(\xx)\DD_{\xxstar} \Xi(\xx, \xxstar)|_{\xxstar = \DD S(\xx)} = 0, \quad \forall \xx$. 
\end{enumerate}
It holds that \eqref{A} $\implies$ \eqref{B} $\implies$ \eqref{C}, but not the other way around. Option \eqref{C} is the least restrictive, as it only implies the energy conservation within gradient flows. Option \eqref{B} is needed for the thermodynamic potential $\Phi(\xx) = -S(\xx) + T_0^{-1}E(\xx)$ (where $T_0$ is the equilibrium temperature of the system) to be a generating potential, as required in the contact-geometric formulation of GENERIC \cite{grmela2010multiscale, pavelka2018multiscale, esen2022role}. Option \eqref{A} is advocated in the context of large deviations \cite{kraaij2020fluctuation} and is useful for adapting constraints irrespective of the concrete form of the entropy. This most restrictive formulation guarantees that the rate function of the large deviation principle is infinite in case of violation of the energy conservation (i.e., such events must occur with zero probability). An infinitesimal formulation of \emph{(A)}, assuming sufficient smoothness, is given in \cite{grmela2018generic} as $\DD E^T(\xx)\DD_{\xxstar} \Xi(\xx, \xxstar) = 0, \ \forall \xx, \xxstar$.
For this paper, we impose \eqref{A} in view of its statistical mechanics foundation and because it can be done via a simple projection, as discussed in Section \ref{sec: nn-generic}.

\section{Nonlinear GENERIC-Embedded Neural Networks (N-GENNs) \label{sec: nn-generic}\label{sec:fginns}}
\begin{figure}[htbp]
    \centering
    \includegraphics[width=1.0\linewidth]{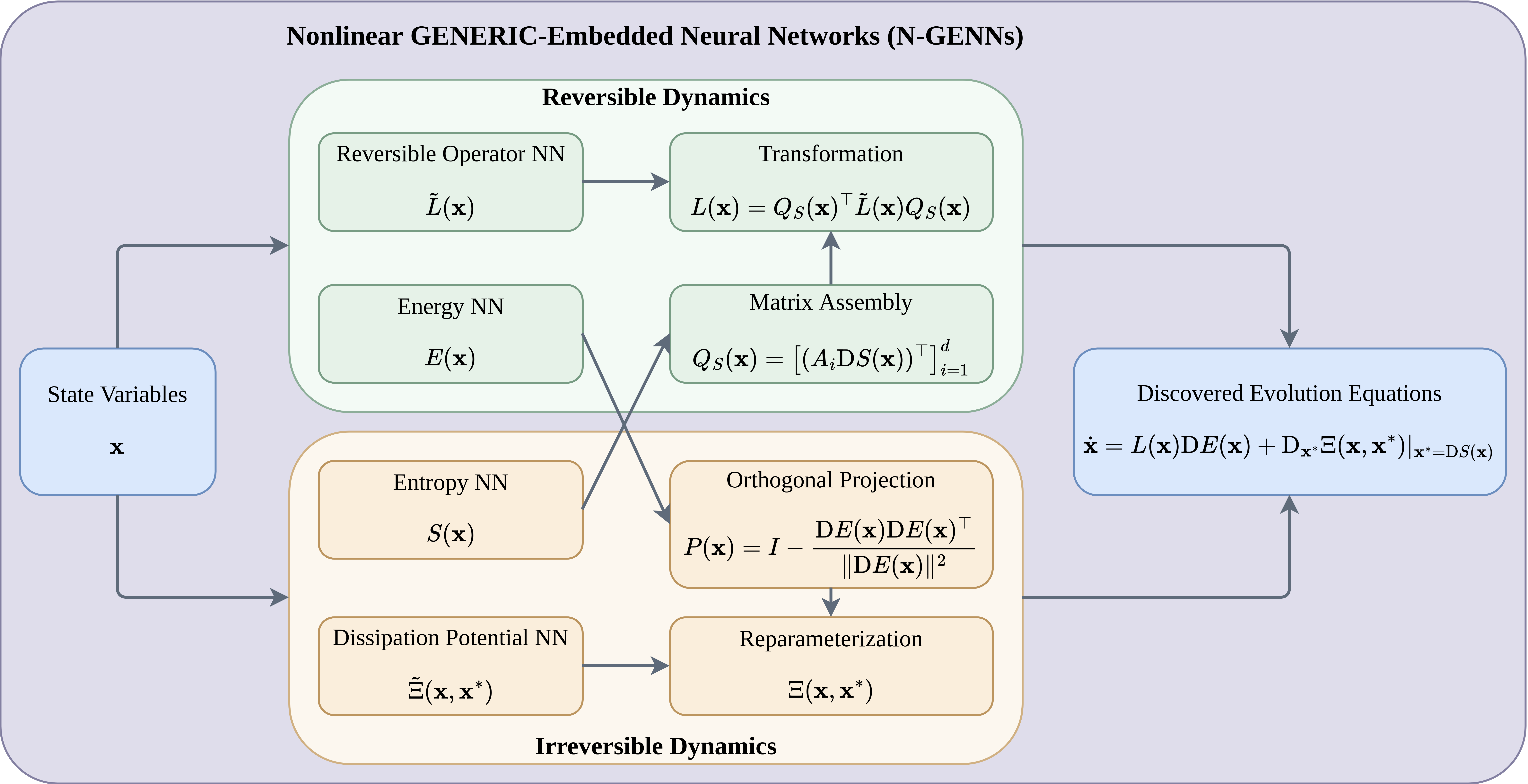}
    \caption{Schematic of N-GENNs. From the state variables $\xx$, the framework learns the reversible operator network $\tilde{L}(\xx)$, the energy network $E(\xx)$, the entropy network $S(\xx)$, and the dissipation-potential network $\tilde{\Xi}(\xx,\xxstar)$; these outputs are then combined through the matrix-assembly block $Q_S(\xx)$, the transformation $L(\xx)=Q_S(\xx)^T\tilde{L}(\xx)Q_S(\xx)$, the orthogonal projection $P(\xx)$, and the irreversible reparameterization $\Xi(\xx,\xxstar)=\tilde{\Xi}\!\left(\xx,P(\xx)\xxstar\right) - \tilde{\Xi}(\xx,0) - \left.\DD_{\xxstar}\tilde{\Xi}(\xx,\xxstar)\right|_{\xxstar=0} P(\xx)\xxstar$ to obtain the discovered evolution equations. All boxed components in the reversible dynamics and irreversible dynamics blocks are parameterized, but the corresponding parameter notation is omitted in the figure for visual clarity.}
    \label{fig:overview_NGINNs}
\end{figure}

We introduce a framework that couples neural networks designed to learn Hamiltonian dynamics with those tailored for generalized gradient flows, yielding a model capable of discovering nonlinear GENERIC evolution equations; see Fig.~\ref{fig:overview_NGINNs}.

As we said in the introduction, Hamiltonian dynamics have been thoroughly explored, such as in the reversible part of GFINNS \cite{zhang2022gfinns}. Complementarily, a scheme capable of learning the dynamics of generalized gradient flows has been explored through Partially Input Convex Neural Networks (PICNNs) in Variational Onsager Neural Networks (VONNs) \cite{huang2022variational} and its dual formulation, such as in \cite{qiu2025bridging}.

This leads to the simple option of separating the deep learning of the non-linear GENERIC building blocks into the following components: scalar neural networks to approximate the thermodynamic energy and entropy, a PICNN to model the dissipation potential, and a reparameterized bivector $L$, as in GFINNs \cite{zhang2022gfinns}. The nontrivial problem that arises is the implementation of the degeneracy conditions \eqref{constr:1} and \eqref{constr:2}.

The enforcement of the condition \eqref{constr:1} was already solved in GFINNS \cite{zhang2022gfinns}, which we will follow for this part. Let the reversible operator generated naively by the neural networks be $\tilde L(\xx)$. We first randomly initialize a vector of skew-symmetric matrices $A \in \mathbb{R}^{d\times d\times d}$:
\begin{equation}
    A = \begin{pmatrix}
        A_1 \\
        \vdots \\
        A_d
    \end{pmatrix}
\end{equation}
where $d$ is the dimension of $\xx$. From this vector, we construct the matrix $Q_S\in \mathbb{R}^{d\times d}$ as
\begin{equation}
    Q_S = \begin{pmatrix}
        (A_1 \DD S)^T \\
        (A_2 \DD S)^T \\
        \vdots \\
        (A_d \DD S)^T
    \end{pmatrix}.
\end{equation}
And finally, we introduce the new reversible operator to be $L(\xx) = Q^T_S(\xx) \tilde L(\xx) Q_S(\xx)$, which naturally inherits the skew-symmetry of $\tilde L(\xx)$. Furthermore, it simultaneously holds that $L(\xx) \DD S = 0$, since $(Q_S(\xx) \DD S)_i = \DD S^T A^T_i \DD S=0$ thanks to the skew-symmetry of $A_i$. Furthermore, in \cite{zhang2022gfinns}, the authors prove the sufficient expressivity of this design.

Now we come to \eqref{constr:2}. Let $P$ be the projection matrix to the orthogonal subspace of $\DD E$, $P(\xx) = I - \DD E\, \DD E^T/||\DD E||^2$. Denoting the raw output of a PICNN as $\tilde\Xi(\xx,\xx^*)$, we define the dissipation potential used in the GENERIC dynamics as
\begin{equation}
    \Xi(\xx,\xxstar)
    =
    \tilde{\Xi}\!\left(\xx,P(\xx)\xxstar\right)
    - \tilde{\Xi}(\xx,0)
    -  {(\xxstar)}^T P(\xx)
        \DD_{\xxstar}\tilde{\Xi}(\xx,\xxstar)|_{\xxstar=0}
    \label{eq:xi_reparam}
\end{equation}
This reparameterization preserves the convexity in $\xxstar$, thereby ensuring property (iv), since it only subtracts an affine function and the composition of a linear map and a convex map is also convex. Moreover, it enforces $\Xi(\xx,0)=0$ and $\DD_{\xxstar}\Xi(\xx,0)=0$, so that $\xxstar=0$ is a global minimizer and $\Xi(\xx,\xxstar) \geq 0$, which guarantees (iii). Finally, by construction, $P(\xx)\DD E(\xx)=0$, and thus (ii) is satisfied. We note that orthogonal projections have been previously used in the context of linear GENERIC to enforce degeneracy conditions \cite{gruber2025efficiently}.

For a schematic of N-GENNs and how the different building blocks are integrated, we refer the reader to Fig.~\ref{fig:overview_NGINNs}. The general loss function $\mathcal{L}$ of N-GENNs reads
\begin{equation}\label{eq:loss}
    \mathcal{L} = \frac{1}{N} \sum^N_{i = 1} \lVert\dot{\xx_i}^{\mathrm{(data)}} - L^{\mathrm{(NN)}}(\xx_i)\DD E^{\mathrm{(NN)}}(\xx_i) - \DD_{\xxstar} \Xi^{\mathrm{(NN)}}(\xx_i,\xxstar)|_{\xxstar = \DD S^{\mathrm{(NN)}}(\xx_i)} \rVert^2,  
\end{equation}
where the superscript (NN) denotes the outputs of the neural networks whereas (data) denotes the corresponding ground-truth data. Furthermore, in Eq.~\eqref{eq:loss} $\DD$ represents the derivative obtained using automatic differentiation. Finally, $N$ is the training batch size and the subscript $i$ indexes the individual samples within the batch.

\paragraph{Non-uniqueness}
As already noted in the introduction, the identification of GENERIC building blocks from macroscopic data is inherently non-unique. This non-uniqueness has been demonstrated in several contexts. From a dynamical perspective, chemical reaction systems can be equivalently described using either quadratic or non-quadratic dissipation potentials \cite{hutter2013quasi}. Similarly, analytical examples for diffusion processes show that multiple pairs of dissipative matrices and thermodynamic potentials can give rise to identical dynamics \cite{huang2022variational}.

In light of this, we evaluate the learned models primarily at the level of the induced dynamics, obtained via automatic differentiation of the identified GENERIC components, rather than through direct comparison of the individual building blocks with reference expressions.

\paragraph{Jacobi identity}
As noted above, the Jacobi identity (\ref{constr:6}) is commonly imposed as one of the defining constraints of GENERIC evolutions. Importantly, however, the Jacobi identity does not directly influence the satisfaction of the thermodynamic laws. Instead, its role is structural: it ensures that the reversible dynamics defines a genuine Poisson bracket, thereby endowing the state space with a consistent geometric interpretation. In particular, the Jacobi identity guarantees time-structure invariance, coordinate independence, and compatibility with the Hamiltonian structure underlying microscopic deterministic dynamics, where the canonical Poisson operator trivially satisfies the Jacobi identity. From a multiscale perspective, this condition underpins the correspondence between stochastic mesoscopic or macroscopic descriptions and their Hamiltonian deterministic counterparts \cite{edwards1997time,pavelka2018multiscale,grmela2026rheology}.

From a practical standpoint, this requirement poses a significant challenge for data-driven modeling. Standard function approximators, such as neural networks, do not generically satisfy the Jacobi identity. In three-dimensional state spaces, special constructions exist in which neural networks can be designed to fulfill the Jacobi identity exactly; however, in higher dimensions the identity is typically enforced only in a weak sense, for example by adding a squared Jacobi residual to the loss function \cite{vsipka2023direct}. To the best of our knowledge, there is currently no general neural network architecture that enforces the Jacobi identity strongly by construction in high-dimensional settings \cite{gruber2025efficiently, urdeitx2025comparison}, as doing so effectively requires solving the Jacobi identity as a system of partial differential equations for the Poisson structure itself \cite{essen2017jacobi}. In light of these considerations, and because the Jacobi identity is not required for thermodynamic admissibility, we do not explicitly assess its exact fulfillment in the examples considered in the remainder of this paper.

\section{Numerical examples}\label{sec:numerical-examples}
\subsection{Example 1. Harmonic oscillator in a heat bath}
\label{sec:harmonic_oscillator}

As a first example, we consider a prototypical GENERIC system with a quadratic dissipation potential, which has been widely studied in the literature \cite{kraaij2020fluctuation}: a harmonic oscillator in a heat bath. Although the proposed framework is designed to handle general, potentially non-quadratic dissipation mechanisms, this example serves as a baseline validation, demonstrating that it can accurately recover classical quadratic dissipation potentials just as well.

\subsubsection{Model description}
\begin{figure}[htbp]
    \centering
    \includegraphics[width=0.5\linewidth]{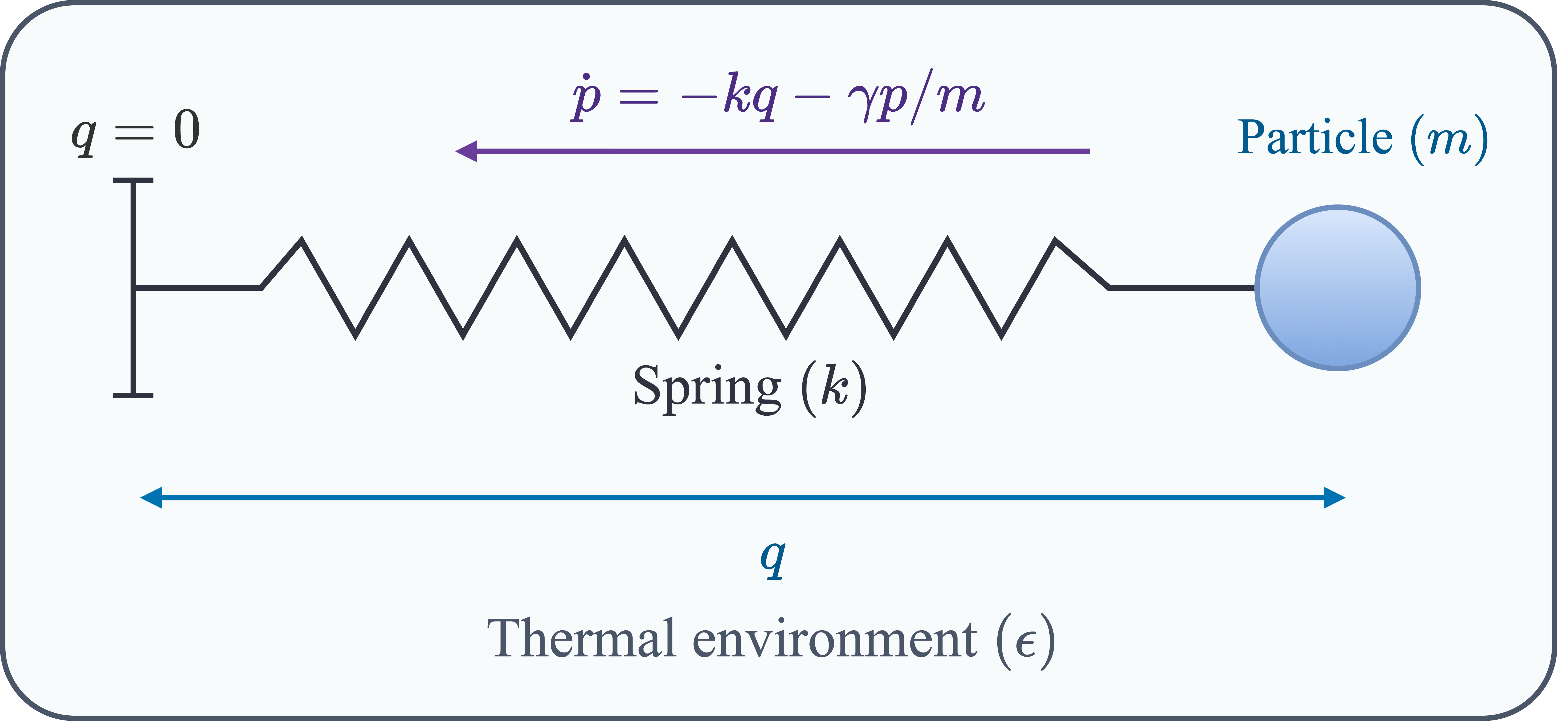}
    \caption{Schematic of the harmonic oscillator in a heat bath (Example 1). A particle of mass $m$ is attached to a linear spring $k$ within a thermal environment ($\epsilon$). Both forces influence the momentum evolution $\dot{p}$, coupling the mechanical energy to the heat bath.}
    \label{fig:harmonic_oscillator_bath_1D}
\end{figure}

A classical harmonic oscillator is characterized by two degrees of freedom representing its position $q$ and conjugate momentum $p$; see Fig.~\ref{fig:harmonic_oscillator_bath_1D}. To describe the oscillator in contact with a heat bath as a closed system whose total energy is conserved, we add a variable $\epsilon$ representing the internal energy of the heat bath. The full state vector is thus given by $\xx = (q, p, \epsilon)$.

The reversible part of the dynamics can be described with the standard Hamiltonian and Poisson operator of the form
\begin{equation}
    E = \frac{p^2}{2m} + \frac{1}{2}kq^2 + \epsilon, \quad L = \begin{pmatrix}
        0 & 1 & 0 \\
        -1 & 0 & 0 \\
        0 & 0 & 0
    \end{pmatrix},
\end{equation}
where $m$ denotes the mass of the oscillator and $k$ is the spring constant of a linear spring connecting the mass to a fixed wall. The dissipative part can then be described using a quadratic dissipation potential $\Xi = \frac{1}{2}(\xx^*)^T M(\xx) \xx^*$ with damping constant $\gamma$, and entropy $S$ as 
\begin{equation}
     M = \gamma T\begin{pmatrix}
        0 & 0 & 0 \\
        0 & 1 & -\frac{p}{m} \\
        0 & -\frac{p}{m} & \frac{p^2}{m^2}
    \end{pmatrix}, \quad S = \frac{\epsilon}{T},
\end{equation}
where $T$ denotes the (constant) temperature of the heat bath. This yields the evolution equations
\begin{equation}
    \dot{q} = \frac{p}{m}, \qquad
    \dot{p} = -kq - \gamma\frac{p}{m}, \qquad
    \dot{\epsilon} = \gamma\frac{p^2}{m^2}.
\end{equation}

\subsubsection{Data generation}
To generate the dataset for this example, we numerically integrate the system from sampled initial conditions using the implicit midpoint method available in standard Python numerical libraries. This method has been utilized multiple times as a benchmark for analogous systems \cite{ottinger2018generic, shang2020structure}. It strictly conserves linear and quadratic invariants of the evolution equations \cite{hairer2006geometric}, ensuring that the first law of thermodynamics is satisfied. Although numerical instabilities can lead to violations of the second law  \cite{betsch2020generic}, the time step used in the examples considered is sufficiently small to prevent such instabilities from occurring. For advances in thermodynamically consistent integrators, we refer the reader to \cite{romero2009thermodynamically, romero2010algorithms, betsch2019energy, scheibl2021structure}.

Initial conditions are sampled uniformly according to $q(0) \in [-1.0, 1.0]$, $p(0) \in [-1.0, 1.0]$ and $\epsilon(0) = 0$, and Table~\ref{tab:oscillator_parameters} documents all simulation parameters that were used in the code. The generated dataset contains 100 trajectories, which are split into 70\% for training, 10\% for validation, and 20\% for testing.
\begin{table}
    \centering
    \caption{Simulation parameters for Example 1, including the dataset size, temporal discretization, and physical parameters of the harmonic oscillator in a heat bath.}
    \label{tab:oscillator_parameters}
    \begin{tabular}{lr}
        \toprule
        \textbf{Parameter} & \textbf{Value} \\
        \midrule
        Trajectories & 100 \\
        Points per trajectory & 1000 \\
        Time step ($\Delta t$) & 0.015 \\
        \midrule
        Oscillator mass ($m$) & 1.0 \\
        Spring stiffness ($k$) & $1.2$ \\
        Damping coefficient ($\gamma$) & 0.4 \\

        \bottomrule
    \end{tabular}
\end{table}

\subsubsection{Results and discussion}
\begin{figure}[htbp]
    \centering
    \includegraphics[width=1.0\linewidth]{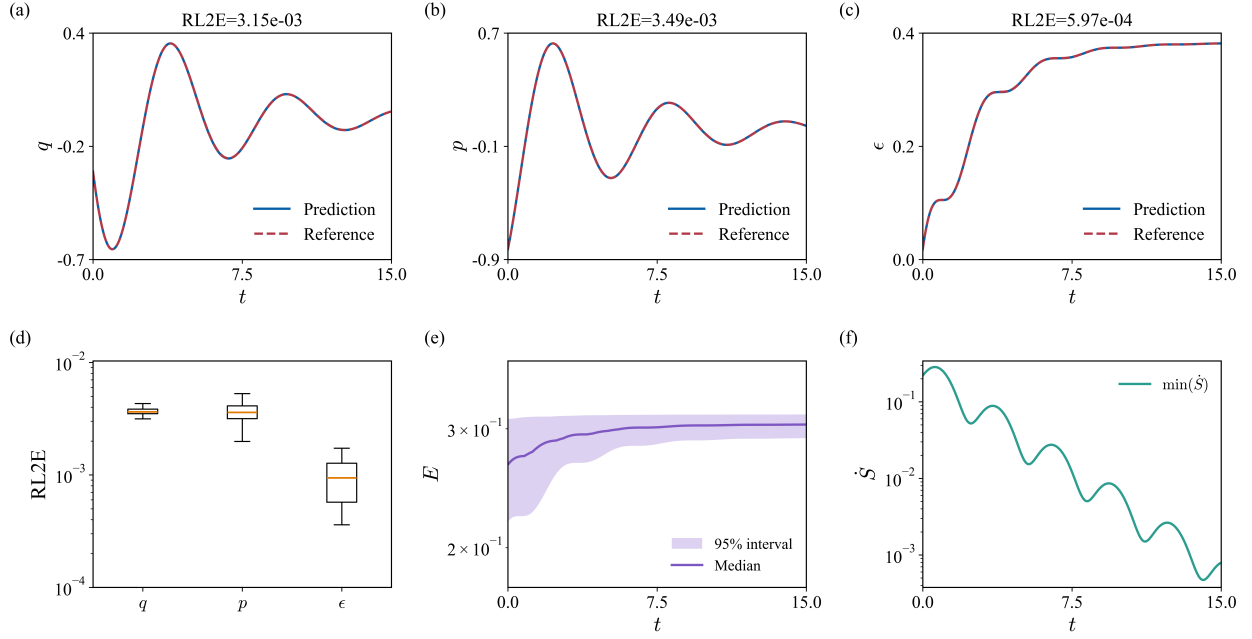}
    \caption{Numerical results for the harmonic oscillator in a heat bath. (a)--(c) Comparison of the learned GENERIC prediction (solid) with the reference trajectory (dashed) for a representative test trajectory: (a) position $q$, (b) momentum $p$, (c) internal energy of the heat bath $\epsilon$; the trajectory-wise relative $\ell^2$ error (RL2E) is reported in each panel. (d) Boxplots of the RL2E for all state variables across the full test dataset. (e) Time evolution of the learned total energy $E$ across the test dataset, showing the median and 95\% interval. (f) Minimum entropy production, $\min(\dot{S})$, across the test dataset at each time step.}
    \label{fig:example1_result_plot}
\end{figure}

Fig.~\ref{fig:example1_result_plot} summarizes the predictive performance of the GENERIC model on both a representative rollout (panels~(a)--(c)) and the full test set (panels (d)--(f)). Panels~(a)--(c) show close agreement between the predicted and reference trajectories for all three state variables. The trajectory-wise relative $\ell^2$ errors for such example are $3.56\times10^{-3}$ for $q$, $3.81\times10^{-3}$ for $p$, and $8.49\times10^{-3}$ for $\epsilon$. These errors confirm that the model captures both the oscillatory energy exchange between $q$ and $p$ and the monotonic increase of internal energy $\epsilon$ due to dissipation. The boxplots in panel~(d) show that these error levels are representative of the entire test set: the medians for all state variables are of order $10^{-3}$. All three distributions exhibit compact interquartile ranges and indicate that the model generalizes well across different initial conditions.

We note that all rollout predictions in this example are produced using a standard fourth-order Runge--Kutta (RK4) integrator, which does not preserve the GENERIC structure at the discrete level. This choice is intentional, as it provides a stringent test of whether the learned dynamics retain the correct thermodynamic structure even when integrated with an accurate, but non-structure-preserving scheme.

Evidence of this robustness is shown in panels~(e) and~(f). Panel~(e) shows the learned total energy $E$ across the test dataset. After a short initial transient, both the median and the 95\% interval remain nearly constant over the time horizon. Panel~(f) displays the minimum entropy production, $\min(\dot{S})$, computed at each time step across the test dataset. This quantity remains strictly positive at all times, thereby confirming consistency with the second law of thermodynamics. 

Overall, this example featuring a quadratic dissipation potential and linear irreversible dynamics serves as a baseline validation of the proposed approach, before proceeding to more challenging settings in the following examples. Further details about the training, rollout prediction, and neural-network architectures can be found in~\ref{sec:nn_train_details}.

\subsection{Example 2. Idealized chemical motor}
\label{sec:chemical_motor}
To further assess the capabilities of the proposed architecture, we consider an idealized chemical motor, which converts chemical energy into mechanical motion. The system of equations is described by a non-canonical Poisson bracket (piston motion) and a non-quadratic dissipation potential (chemical reaction), providing a sufficiently challenging non-linear reversible-irreversible coupling \cite{pavelka2018multiscale}.

\subsubsection{Model description}
\begin{figure}[htbp]
    \centering
    \includegraphics[width=0.6\linewidth]{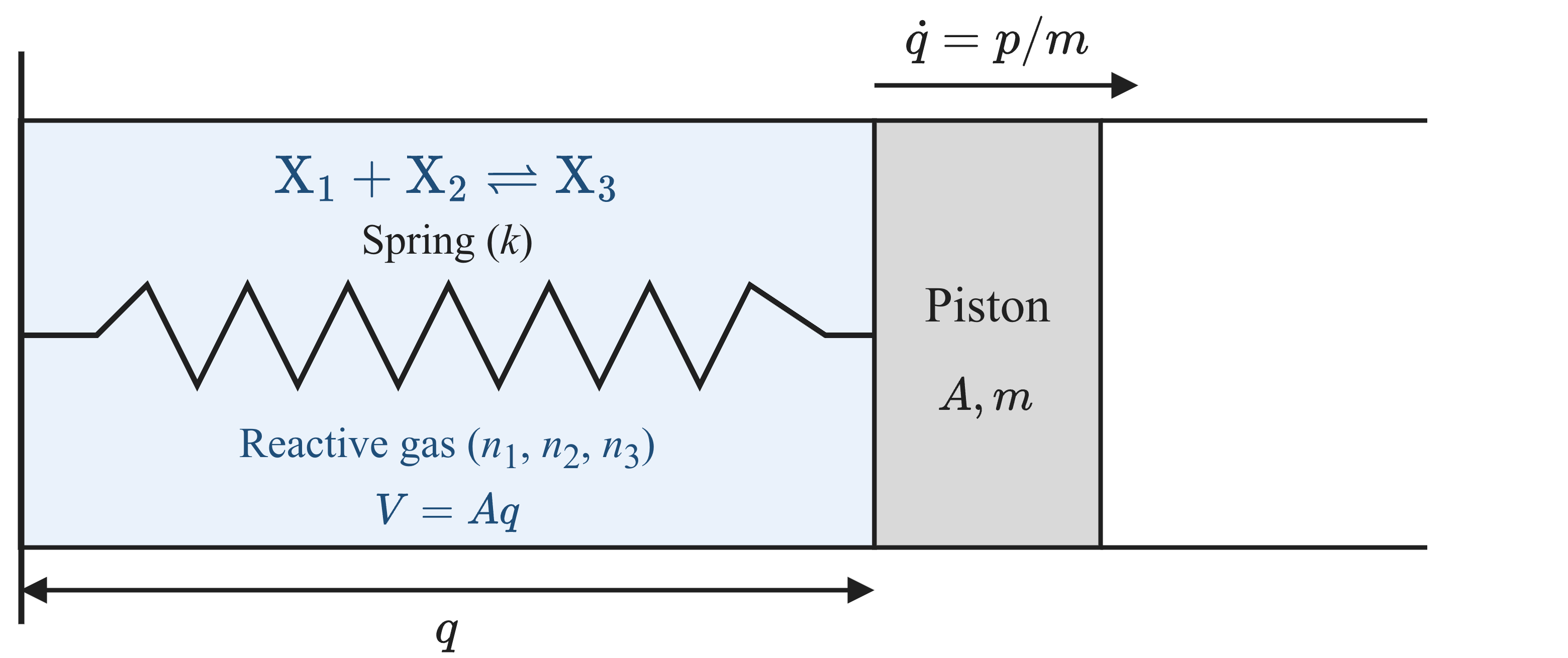}
    \caption{Schematic of the idealized chemical motor (Example~2). A spring attached to the piston couples mechanical motion $(q,p)$ to a reactive van der Waals gas mixture $(n_1,n_2,n_3,\epsilon)$ undergoing $\mathrm{X}_1+\mathrm{X}_2\rightleftharpoons \mathrm{X}_3$. The gas chamber volume is denoted by $V$. The gas pressure generates a reversible force $F(\mathbf{x})$ on the piston, while reaction kinetics are governed by a non-quadratic dissipation potential $\Xi$.}
    \label{fig:ideal_motor}
\end{figure}

We consider an adiabatically isolated piston-spring system, where the piston has mass $m$, is attached to a linear spring with stiffness $k$, and has cross-sectional area $A$; see Fig.~\ref{fig:ideal_motor}. This piston contains a mixture of three chemical species $\mathrm{X}_1$, $\mathrm{X}_2$, and $\mathrm{X}_3$, modeled as real van der Waals gases, that undergo a chemical reaction $\mathrm{X}_1 + \mathrm{X}_2 \rightleftharpoons \mathrm{X}_3$.

The state variables for this problem then are $\xx = (q,p,\epsilon,n_1,n_2,n_3)$ where $q$ and $p$ are the position and momentum of the piston ($q$ is intrinsically tied to the volume of the piston through $V = Aq$), $n_\mu$ are the number of moles for each species, and $\epsilon$ is their combined internal energy.

The Hamiltonian for this problem accounts for the kinetic energy of the piston, the potential energy of the spring, and the internal energy of the three gases. If the entropy of the gases $s$ were among the state variables, instead of the internal energy $\epsilon$, the Poisson bracket would be canonical. When $\epsilon$ replaces the entropy, the interaction force between the gases and the piston appears in the reversible operator as momentum--energy coupling,
\begin{equation}\label{eq:motor_energy}
    E = \frac{p^2}{2m} + \frac{1}{2}k(q-q_0)^2 + \epsilon, \quad L = \begin{pmatrix}
        0 & 1 & 0 & 0 & 0 & 0 \\
        -1 & 0 & F(\xx) & 0 & 0 & 0 \\
        0 & -F(\xx) & 0 & 0 & 0 & 0 \\
        0 & 0 & 0 & 0 & 0 & 0 \\
        0 & 0 & 0 & 0 & 0 & 0 \\
        0 & 0 & 0 & 0 & 0 & 0 
    \end{pmatrix}
\end{equation}
where $F(\xx) = \left(\frac{\partial S}{\partial q}\right)_{p,\epsilon}/\left(\frac{\partial S}{\partial \epsilon}\right)_{p,q}$.
This leads to the equations of motion for the first three state variables (governed by $L\DD E$ exclusively)
\begin{equation}
    \dot{q} = \frac{p}{m}, \quad \dot{p} = -k(q-q_0) + F(\xx), \quad \dot{\epsilon} = -F(\xx)\frac{p}{m}.
\end{equation}
To find the force $F(\xx)$, we need to specify the energy of the system or the entropy as a function of energy and volume (and thus also of position $q$). 
For the entropy, we use the Sackur--Tetrode formula modified for van der Waals gases \cite{callen1960thermodynamics}, adding the mean-field van der Waals corrections to the volume and internal energy. We work with entropy scaled by the gas constant, i.e., $S\equiv S_{\mathrm{phys}}/R$, so that thermodynamic forces in the dissipation potential remain dimensionless.
\begin{equation}
    S(\xx) = \sum_\alpha n_\alpha\bigg[\frac{5}{2} + \ln\left(\frac{(V - \sum_\nu n_\nu b_\nu)}{n_\alpha N_A }\left(\frac{4\pi m_\alpha \left(\epsilon + \frac{\sum_{\rho\sigma}n_\rho n_\sigma a_{\rho\sigma}}{V}\right)}{3h^2N_A\sum_\beta n_\beta}\right)^{\frac{3}{2}}\right)\bigg].
\end{equation}
In this formula, $m_\mu$ represents the masses of individual molecules, $N_A$ is the Avogadro constant, $h$ is the Planck constant, parameters $b_\mu$ represent the volume that the gas molecules occupy, and $a_{\mu \nu}$ represents the interaction energy between these molecules. In SI units, this corresponds to $[m_\mu]=\mathrm{kg}$, $[b_\mu]=\mathrm{m^3\,mol^{-1}}$, $[a_{\mu\nu}]=\mathrm{Pa\,m^6\,mol^{-2}}$, and $[\epsilon]=\mathrm{J}$. 
With $n_\mu$ interpreted as moles, this implies the van der Waals relations
\begin{equation}
    \frac{F}{A}=P = \frac{nRT}{V - \sum_\nu n_\nu b_\nu} - \frac{\sum_{\rho\sigma}n_\rho n_\sigma a_{\rho\sigma}}{V^2},
    \quad
    \epsilon = \frac{3}{2}nRT - \frac{\sum_{\rho\sigma}n_\rho n_\sigma a_{\rho\sigma}}{V},
\end{equation}
where $n = \sum_\beta n_\beta$.

For the remainder of this example, we set $R = h = N_A = 1$ and omit these constants from the formulas below.

The chemical reaction is described by the classic law of mass action, with dissipation potential in the form \cite{grmela2010multiscale}
\begin{equation}
    \Xi(\xx, \xxstar) = \alpha\sqrt{n_1 n_2 n_3}\bigg[\cosh\left(\frac{n^*_1 + n^*_2 - n^*_3}{2}\right)-1\bigg]
\end{equation}
where $\alpha > 0$ is the rate coefficient, and $n_\alpha^*$ are the conjugate numbers of moles (eventually identified with the corresponding derivatives of entropy). Note that with this combination of the Hamiltonian and the dissipation potential, the degeneracy condition \eqref{constr:1} is immediately satisfied.

Through calculation of derivatives and using the van der Waals equations of state, we can then find the dynamics of the molar concentrations $n_\mu$ and the precise formula for the force $F(\xx)$. In detail, this derivation is done in \ref{sec:motor_derivation}. The evolution equations then are
\begin{align}
    \dot{q} &= \frac{p}{m}, \nonumber\\
    \dot{p} &= -k(q-q_0) + \frac{2}{3}\frac{\epsilon Aq + \sum_{\rho\sigma}n_\rho n_\sigma a_{\rho\sigma}}{q(Aq - \sum_\nu n_\nu b_\nu)} - \frac{\sum_{\rho\sigma}n_\rho n_\sigma a_{\rho\sigma}}{Aq^2},\nonumber\\
    \dot{\epsilon} &= -\frac{p}{m}\left(\frac{2}{3}\frac{\epsilon Aq + \sum_{\rho\sigma}n_\rho n_\sigma a_{\rho\sigma}}{q(Aq - \sum_\nu n_\nu b_\nu)} - \frac{\sum_{\rho\sigma}n_\rho n_\sigma a_{\rho\sigma}}{Aq^2}\right), \nonumber\\
    \dot{n}_1 &= \frac{\alpha}{4}(Kn_3 - \frac{1}{K}n_1 n_2) ,\\
    \dot{n}_2 &= \frac{\alpha}{4}(Kn_3 - \frac{1}{K}n_1 n_2) ,\nonumber\\
    \dot{n}_3 &= \frac{\alpha}{4}(\frac{1}{K}n_1 n_2 - Kn_3), \nonumber
\end{align}
where $K$ is the reaction constant
\begin{equation}
    K = \exp\bigg[\frac{3n \sum_{\sigma} (a_{1\sigma} + a_{2\sigma} - a_{3\sigma})n_\sigma }{2(\epsilon Aq + \nn^T \aaa \nn)} - \frac{n(b_1 + b_2 - b_3)}{2(Aq - \nn^T \bb)}\bigg]\left(\frac{m_1 m_2}{m_3}\right)^{\frac{3}{4}} (Aq - \nn^T\bb)^{\frac{1}{2}}\left(\frac{4\pi (\epsilon + \frac{\nn^T\aaa\nn}{Aq})}{3n}\right)^{\frac{3}{4}}.
\end{equation}
In \ref{sec:motor_derivation}, we further show that this model satisfies the second degeneracy condition and the Jacobi identity.
\subsubsection{Data generation}
The trajectories were generated using the implicit midpoint integration method. Since the total energy is again a quadratic invariant of the state variables (see Eq.~\eqref{eq:motor_energy}), the situation is analogous to Example 1. For numerical stability, the trajectories are generated in the dimensionless variables introduced above, with $R = h = N_A = 1$.  Table \ref{tab:motor_parameters} then documents all simulation parameters.

The initial conditions for trajectories were uniformly sampled as follows: the initial molar concentrations satisfy $n_1(0) \in [0.8, 1.3]$, $n_2(0) \in [1.2, 2.4]$, and $n_3(0) \in [0.002, 0.003]$. The initial momentum was fixed as $p(0) = 0$ for all trajectories, while the initial internal energy and initial position are sampled as $\epsilon(0) \in [2.4, 6.7]$ and $q(0) \in [1.2, 2.8]$, respectively. In total, 200 trajectories are generated, of which 70\% are used for training, 10\% for validation, and 20\% for testing.

\begin{table}[H]
    \centering
    \caption{Simulation parameters for Example 2, including the dataset size, temporal discretization, and physical parameters of the idealized chemical motor.}
    \label{tab:motor_parameters}
    \begin{tabular}{lr}
        \toprule
        \textbf{Parameter} & \textbf{Value} \\
        \midrule
        Trajectories & 200 \\
        Points per trajectory & 1000 \\
        Time step ($\Delta t$) & 0.01 \\
        \midrule
        Piston mass ($m$) & 0.8 \\
        Piston area ($A$) & 1.0 \\
        Spring constant ($k$) & $31.6$ \\
        Natural spring length ($q_0$) & 2.0 \\
        Rate factor ($\alpha$) & 2.0 \\
        \midrule
        Matrix $a$ & $\begin{pmatrix} 0.1 & 0.1 & 0.1 \\ 0.1 & 0.1 & 0.1 \\ 0.1 & 0.1 & 20.0 \end{pmatrix}$ \\
        \addlinespace[5pt]
        Vector $b$ & $\begin{pmatrix} 0.05 & 0.05 & 0.05 \end{pmatrix}^T$ \\
        \midrule
        $m_1$, $m_2$, $m_3$ & 1.0, 1.2, 2.2 \\
        \bottomrule
    \end{tabular}
\end{table}

\subsubsection{Results and discussion}
\begin{figure}[htbp]
    \centering
    \includegraphics[width=1.0\linewidth]{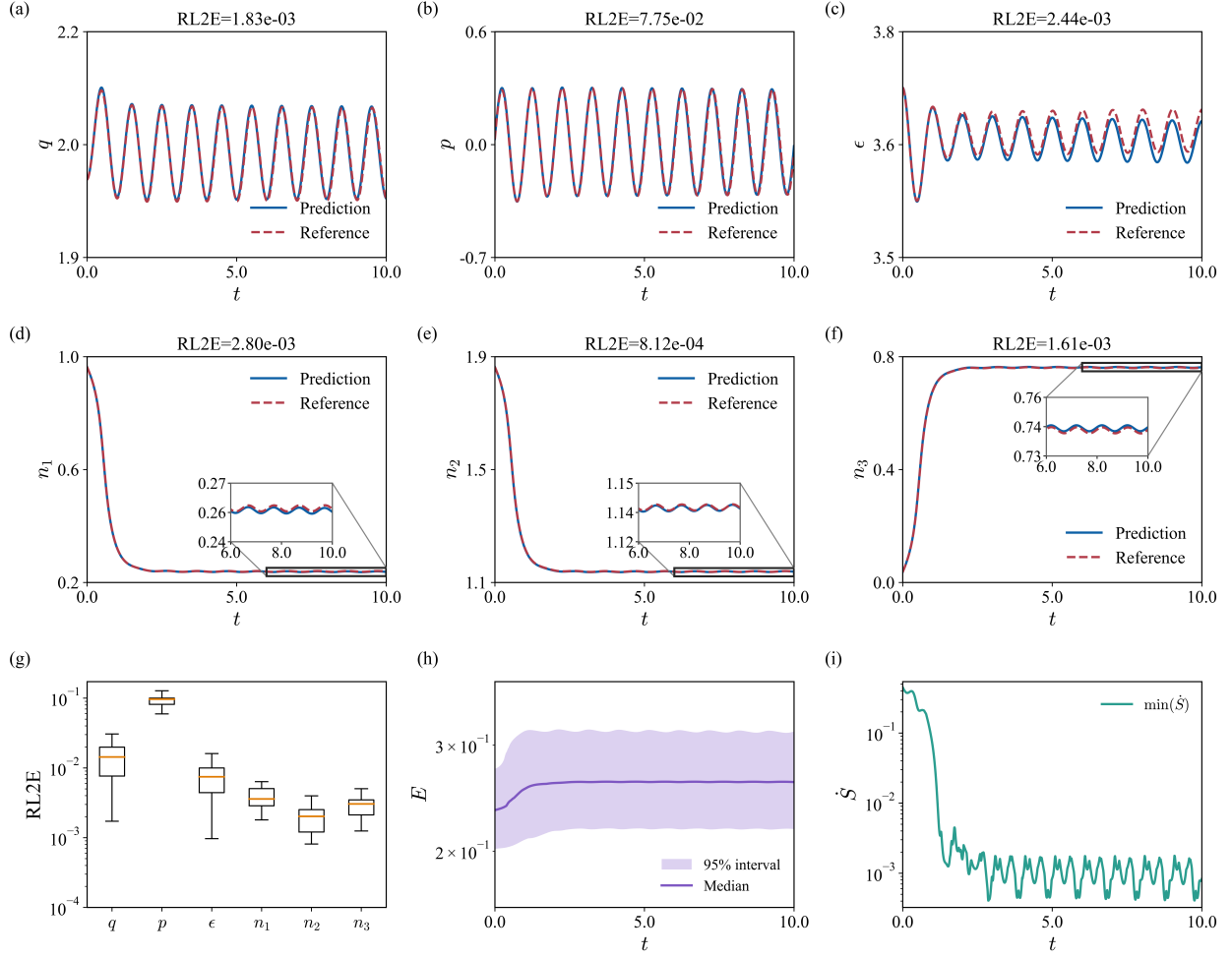}
    \caption{Numerical results for the idealized chemical motor. (a)--(f) Comparison of the learned GENERIC prediction (solid) with the reference trajectory (dashed) for a representative test trajectory: (a) position $q$, (b) momentum $p$, (c) internal energy $\epsilon$, (d)--(f) mole numbers $n_1$, $n_2$, $n_3$; the trajectory-wise relative $\ell^2$ error (RL2E) is reported in each panel. (g) Boxplots of the RL2E for all state variables across the full test dataset. (h) Time evolution of the learned total energy $E$ across the test dataset, showing the median and 95\% interval. (i) Minimum entropy production, $\min(\dot{S})$, across the test dataset at each time step.}
    \label{fig:example2_result_plot}
\end{figure}

Compared with the harmonic oscillator in Section~\ref{sec:harmonic_oscillator}, this system exhibits stronger nonlinear dynamics due to the van der Waals equation of state and the chemical reaction kinetics, which is governed by a non-quadratic dissipation potential. Fig.~\ref{fig:example2_result_plot} demonstrates the predictive performance of the GENERIC model on this six-dimensional system. Panels~(a)--(f) show close agreement between the predicted and reference trajectories for all six state variables, with trajectory-wise relative $\ell^2$ errors of $1.17\times10^{-3}$ for $q$, $5.99\times10^{-2}$ for $p$, $1.49\times10^{-3}$ for $\epsilon$, $1.49\times10^{-3}$ for $n_{1}$, $2.16\times10^{-4}$ for $n_{2}$, and $8.24\times10^{-4}$ for $n_{3}$. These results indicate that the model accurately captures both the oscillatory piston mechanics in $(q,p)$ and the reaction-driven evolution of $(n_1,n_2,n_3)$, together with the associated thermodynamic response in $\epsilon$. Panel~(g) further confirms that these error levels hold across the full test set: all six variables maintain low median errors with narrow spreads, even under the strong nonlinear coupling present in this example.

As in Section~\ref{sec:harmonic_oscillator}, rollout predictions are produced with an RK4 integrator, so the thermodynamic diagnostics reported below constitute empirical evidence rather than exact structural guarantees. Panel~(h) shows the learned total energy $E$ across the test dataset. Its median undergoes only a mild initial adjustment and then remains nearly constant, while the 95\% interval stays stable throughout the rollout. Panel~(i) shows the minimum entropy production, $\min(\dot{S})$, computed at each time step across the test dataset, and it remains strictly positive throughout. These diagnostics confirm that, even under stronger reversible--irreversible coupling and a non-quadratic dissipation potential, the architectural constraints built into the network are sufficient to maintain a nearly conserved total energy together with non-negative entropy production. The hyperparameters of the neural-network architectures and training for this example are listed in~\ref{sec:nn_train_details}.

\subsection{Example 3. Viscoplasticity}
\label{sec:viscoplasticity}
As a final example, we consider a continuum system governed by a viscoplastic model of Perzyna type with linear kinematics \cite{lubliner2008plasticity,simo1998computational}, a widely used constitutive framework for describing rate-dependent inelastic behavior in solids. In contrast to the previous examples, which are formulated as systems of ordinary differential equations, this model involves spatially distributed fields, and is hence governed by partial differential equations. Moreover, its dynamics are driven by a non-quadratic dissipation potential associated with the viscoplastic flow. Together, these features make it a relevant example to assess the proposed framework in the context of learning nonlinear irreversible dynamics in a spatially extended setting.

\subsubsection{Model description}
We consider a one-dimensional Perzyna viscoplastic model \cite{lubliner2008plasticity,simo1998computational}. The state vector $\xx = (u, p, \varepsilon^{vp}, \theta )$, is composed of the displacement field $u(x, t)$, momentum $p(x, t)$, viscoplastic strain $\varepsilon^{vp}(x, t)$ and temperature $\theta(x, t)$. Here, $x$ and $t$ are the spatial and temporal coordinates, respectively. The strain is defined as $\varepsilon = \frac{\partial u}{\partial x}$, and the elastic strain is defined as $\varepsilon^e = \varepsilon - \varepsilon^{vp}$. Following the derivation in \cite{mielke2011formulation}, the GENERIC governing equations are governed by the energy functional
\begin{equation}
    E[\xx] = \int \left(\frac{1}{2 \rho} p^2 + c \theta + \frac{1}{2} C (\varepsilon - \varepsilon^{vp})^2 \right)\, \dd x,
\end{equation}
and the entropy functional
\begin{equation}
    S[\xx] = \int c \log \theta\, \dd x,
\end{equation}
where $\rho$ is the density, $c$ is the specific heat, and $C$ is the elastic modulus. The corresponding functional derivatives are
\begin{equation}
    \DD E (\xx) = 
    \begin{pmatrix}
        - \frac{\partial }{\partial x} \left( C(\varepsilon - \varepsilon^{vp}) \right)  \\
        \frac{p}{\rho} \\
        C(\varepsilon^{vp} - \varepsilon) \\
        c \\
    \end{pmatrix} \quad \text{and} \quad
    \DD S (\xx) = 
    \begin{pmatrix}
        0  \\
        0 \\
        0 \\
        \frac{c}{\theta} \\
    \end{pmatrix}.
\end{equation}
The reversible dynamics are generated by the Poisson operator
\begin{equation}
    L = 
    \begin{pmatrix}
        0 & I & 0 & 0 \\
        -I & 0 & 0 & 0 \\
        0 & 0 & 0 & 0 \\
        0 & 0 & 0 & 0 \\
    \end{pmatrix},
\end{equation}
and the dissipation potential is constructed as 
\begin{equation}
    \Xi[\xx; \xxstar] = \int \frac{\theta}{2 \eta} \left \langle \left| \xi_{\varepsilon^{vp}} + \frac{\xi_\theta}{c} \sigma \right| - \frac{\sigma_y}{\theta} \right \rangle^2 \dd x,
\end{equation}
where $\xxstar = (\xi_u, \xi_p, \xi_{\varepsilon^{vp}}, \xi_\theta)$ are the dual variables, $\eta$ is the viscosity, $\sigma = C(\varepsilon - \varepsilon^{vp})$ is the stress, and $\sigma_y >0$ is the constant yield stress. $|\cdot|$ denotes the absolute value, and, for an arbitrary function $f$, $\langle f \rangle$ is defined as
\begin{equation}
    \langle f \rangle = 
    \begin{cases}
			f, & \text{if $f > 0$}\\
            0, & \text{otherwise.}
    \end{cases}.
\end{equation}
$\Xi$ is convex with respect to $\xxstar$ because it is the composition of a convex function $g(\xi) = \frac{\theta^2}{2 \eta} \left \langle \left| \xi  \right| - \frac{\sigma_y}{\theta} \right \rangle^2$ and a linear function $\xi = \xi_{\varepsilon^{vp}} + \frac{\xi_\theta}{c} \sigma$. It is also straightforward to see that $\Xi[\xx; \xxstar=0] = 0$. Note that $\Xi$ is non-quadratic. Its functional derivative is 
\begin{equation}
    \DD_{\xxstar} \Xi[\xx;\xxstar] = \frac{\theta}{\eta} \left \langle \left|  \xi_{\varepsilon^{vp}} + \frac{\xi_\theta}{c} \sigma\right|-\frac{\sigma_y}{\theta}\right\rangle \text{sign}\left(  \xi_{\varepsilon^{vp}} + \frac{\xi_\theta}{c} \sigma \right) 
    \begin{pmatrix}
        0 \\
        0 \\
        1 \\
        \frac{\sigma}{c} \\
    \end{pmatrix},
\end{equation}
and, therefore, the GENERIC evolution equations of the system read\footnote{For simplicity, the effective heat conductivity is set to be zero. The conductive heat flux could be taken into account if necessary; see section 4.5 of \cite{mielke2011formulation}.}
\begin{align} \label{eq: example 3}
    \begin{pmatrix}
        \dot{u} \\
        \dot{p} \\
        \dot{\varepsilon^{vp}} \\
        \dot{\theta} \\
    \end{pmatrix}
    =
    \begin{pmatrix}
        \frac{p}{\rho} \\
        \frac{\partial }{\partial x} \left
        ( C(\varepsilon - \varepsilon^{vp}) \right)  \\
                \frac{1}{\eta} \langle |\sigma| - \sigma_y \rangle \text{sign}(\sigma) \\
        \frac{\sigma \dot{\varepsilon}^{vp}}{c}
    \end{pmatrix}.
\end{align}

\subsubsection{Data generation}
The spatial domain $[0,l]$ is discretized using a finite element scheme with linear shape functions on a uniform mesh of spacing $\Delta x$. Time integration over $t \in [0,T]$ is performed with a uniform time step $\Delta t$ using a partitioned explicit scheme: the displacement field is advanced with a velocity-Verlet integrator, while the viscoplastic strain and temperature fields are updated using a forward Euler scheme. The velocity-Verlet method is a symplectic time integrator, which provides good long-time stability and ensures that the energy of the reversible (Hamiltonian) part is preserved up to small bounded oscillations, making it well-suited for the mechanical dynamics. The explicit treatment of the dissipative part is adopted for simplicity and computational efficiency; more standard approaches in computational viscoplasticity typically rely on implicit return-mapping algorithms or variational updates to ensure robustness ~\cite{simo1998computational,lubliner2008plasticity}. The time step is chosen as $\Delta t = 0.5 \,\Delta x / \sqrt{C/\rho}$, where $\sqrt{C/\rho}$ denotes the elastic wave speed, to satisfy the Courant–Friedrichs–Lewy (CFL) condition. Details of the discretization scheme are provided in \ref{appx:example3_discretization}.

In the discrete update, the nonsmooth Perzyna flow rule is evaluated pointwise at the element centers. We use the single-valued numerical convention $\operatorname{sign}(0)=0$, and the Macaulay bracket is evaluated as $\langle a\rangle=\max(a, 0)$. Consequently, when $\sigma=0$, or more generally when $|\sigma| \leq \sigma_y$, the computed viscoplastic flow vanishes and no plastic strain increment is applied. This yields a single-valued explicit evolution law suitable for time integration, avoiding the need to solve the associated set-valued differential inclusion.

All trajectories start from the same undeformed state at rest, with the state variables initialized as
\begin{equation}
    u(x,0)=0,\qquad p(x,0)=0,\qquad \varepsilon^{vp}(x,0)=0,\qquad \theta(x,0)=298\,\mathrm{K}.
\end{equation}
A fixed Dirichlet boundary condition is imposed at the left boundary, i.e., $u(0,t)=0$, and a displacement-controlled ramp-and-hold boundary condition is imposed at the right boundary, where the displacement is increased to a prescribed strain $\varepsilon_t$ and subsequently held constant, i.e., $u(l,t)=u_r (t)$, where
\begin{equation}
    u_r(t) = 
    \begin{cases}
			\varepsilon_t l \left( 3 \left(\frac{t}{T_{\text{ramp}}}\right)^2 - 2 \left(\frac{t}{T_{\text{ramp}}}\right)^3 \right), & \text{if $t \leq T_{\text{ramp}}$}\\
            \varepsilon_t l, & \text{otherwise}.
	\end{cases}
\end{equation}
 Thus, all generated trajectories share the same initial conditions and loading protocol, differing only in the value of the target strain $\varepsilon_t$.
 
All parameters, including the dataset size, target strains, spatial and temporal discretization, and material parameters of the viscoplasticity model are summarized in Table \ref{tab:example3_params}.

The generated dataset is partitioned into 70\% for training, 10\% for validation, and 20\% for testing.

\begin{table}[H]
    \centering
    \caption{Simulation parameters for Example 3, including the dataset size, loading protocol, spatial and temporal discretization, and material parameters of the viscoplasticity model.}
    \label{tab:example3_params}
    \begin{tabular}{lr}
        \toprule
        \textbf{Parameter} & \textbf{Value} \\
        \midrule
        Trajectories & 100 \\
        Target strains ($\varepsilon_t$) & Equally spaced in $[1\%,\,2\%]$ \\
        \midrule
        Domain length ($l$) & $0.2\,\mathrm{m}$ \\
        Final time ($T$) & $0.0002\,\mathrm{s}$ \\
        Ramp time ($T_{\mathrm{ramp}}$) & $0.8T$ \\
        \midrule
        Density ($\rho$) & $7800\,\mathrm{kg\cdot m^{-3}}$ \\
        Elastic modulus ($C$) & $210\,\mathrm{GPa}$ \\
        Viscosity ($\eta$) & $8\,\mathrm{GPa\cdot s}$ \\
        Yield stress ($\sigma_y$) & $250\,\mathrm{MPa}$ \\
        Specific heat ($c$) & $1000\,\mathrm{J\cdot kg^{-1}\cdot K^{-1}}$ \\
        \midrule
        Number of elements ($N_x$) & $29$ \\
        Mesh size ($\Delta x$) & $6.90\times10^{-3}\,\mathrm{m}$ \\
        Time step ($\Delta t$) & $6.65\times10^{-7}\,\mathrm{s}$ \\
        \bottomrule
    \end{tabular}
\end{table}

\subsubsection{N-GENNs model}
In this section, we detail the application of the N-GENNs framework to the continuum setting, highlighting the key differences with respect to the previous ODE-based examples. In contrast to finite-dimensional systems, the GENERIC formulation for continuum models is expressed in terms of spatially distributed fields and associated energy and entropy functionals, which are defined as integrals of local densities. Accordingly, we parameterize and learn the corresponding density functions rather than the full functionals themselves.
\begin{equation}
     E[\xx] = \int e \left( \varepsilon, p, \varepsilon^{vp}, \theta \right) \dd x ,
\end{equation}
\begin{equation}
    S[\xx] = \int s(\varepsilon^{vp}, \theta) \dd x,
\end{equation}
\begin{equation}
    \Xi[\xx; \xxstar] = \int \psi(\varepsilon, p, \varepsilon^{vp}, \theta; \xxstar) \dd x.
\end{equation}
In addition, we incorporate problem-specific structural assumptions to reduce the learning complexity and improve physical interpretability. In particular, for this example, we fix the reversible operator to its canonical form, as it is standard for generalized standard materials \cite{mielke2011formulation}, and exploit the known kinetic contribution to the energy density by learning only the remaining part $\omega$ through a neural network representation
\begin{equation}
    e \left( \varepsilon, p, \varepsilon^{vp}, \theta \right) = \frac{p^2}{2 \rho} + \omega \left( \varepsilon, \varepsilon^{vp}, \theta \right).
\end{equation}
Finally, to enforce consistency with the definition of momentum, we require $\frac{\partial \psi}{\partial \xi_{\varepsilon}}=0$, so that $p = \rho \dot{u}$ is automatically guaranteed. 

These choices yield the following functional derivatives of the energy, entropy, and dissipation potential
\begin{equation}
    \DD E
    = 
    \begin{pmatrix}
        - \frac{\partial}{\partial x} \frac{\partial \omega}{\partial \varepsilon} \\
        \frac{p}{\rho} \\
        \frac{\partial \omega}{\partial \varepsilon^{vp}} \\
        \frac{\partial \omega}{\partial \theta} \\
    \end{pmatrix}, \quad
     \DD S = 
    \begin{pmatrix}
        0 \\
        0 \\
        \frac{\partial s}{\partial \varepsilon^{vp}} \\
        \frac{\partial s}{\partial \theta} \\
    \end{pmatrix}, \quad
    \DD_{\xxstar} \Xi[\xx;\DD S] = 
    \begin{pmatrix}
        0 \\
        \frac{\partial \psi}{\partial \xi_p}\Big|_{\xi_p = 0} \\
        \frac{\partial \psi}{\partial \xi_{\varepsilon^{vp}}}\Big|_{\xi_{\varepsilon^{vp}} = \frac{\partial s}{\partial \varepsilon^{vp}}} \\
        \frac{\partial \psi}{\partial \xi_{\theta}}\Big|_{\xi_{\theta} = \frac{\partial s}{\partial \theta}} \\
    \end{pmatrix},
\end{equation}
and, therefore, the following evolution equations
\begin{equation}
    \begin{pmatrix}
        \dot{u} \\
        \dot{p} \\
        \dot{\varepsilon^{vp}} \\
        \dot{\theta} \\
    \end{pmatrix}
    =
    \begin{pmatrix}
        \frac{p}{\rho}  \\
        \frac{\partial}{\partial x} \frac{\partial \omega}{\partial \varepsilon} + \frac{\partial \psi}{\partial \xi_p}\Big|_{\xxstar = \DD S}\\
        \frac{\partial \psi}{\partial \xi_{\varepsilon^{vp}}}\Big|_{\xxstar = \DD S} \\
        \frac{\partial \psi}{\partial \xi_{\theta}}\Big|_{\xxstar = \DD S} \\
    \end{pmatrix}.
\end{equation}

\subsubsection{Results and discussion}
During rollout, the prescribed boundary histories for displacement and momentum are imposed from the reference boundary data, while the interior nodal variables and cell-centered variables are evolved by the learned model. Similar to the previous two examples, an RK4 integrator is used to generate the rollout predictions. Because this example is driven by a prescribed boundary displacement, it is not closed, and hence, its total energy is not conserved.

\begin{figure}[htbp]
    \centering
    \includegraphics[width=1.0\linewidth]{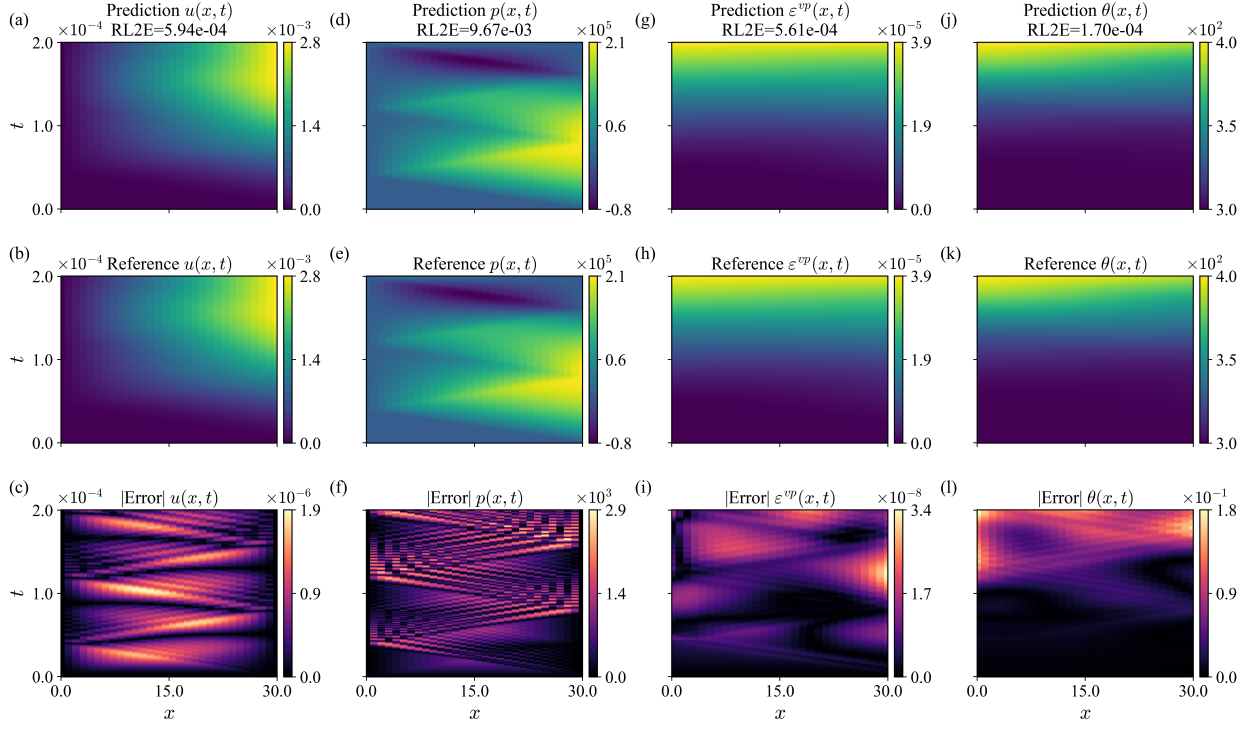}
    \caption{Numerical results for 1D Perzyna viscoplasticity on a representative test trajectory. (a)--(c) Displacement $u(x,t)$, (d)--(f) momentum $p(x,t)$, (g)--(i) viscoplastic strain $\varepsilon^{vp}(x,t)$, (j)--(l) temperature $\theta(x,t)$; within each triplet, the top, middle, and bottom entries show the learned GENERIC prediction, the reference solution, and the pointwise absolute error. The trajectory-wise relative $\ell^2$ error (RL2E) is reported for each field.}
    \label{fig:example3_field_plot}
\end{figure}

\begin{figure}[htbp]
    \centering
    \includegraphics[width=1.0\linewidth]{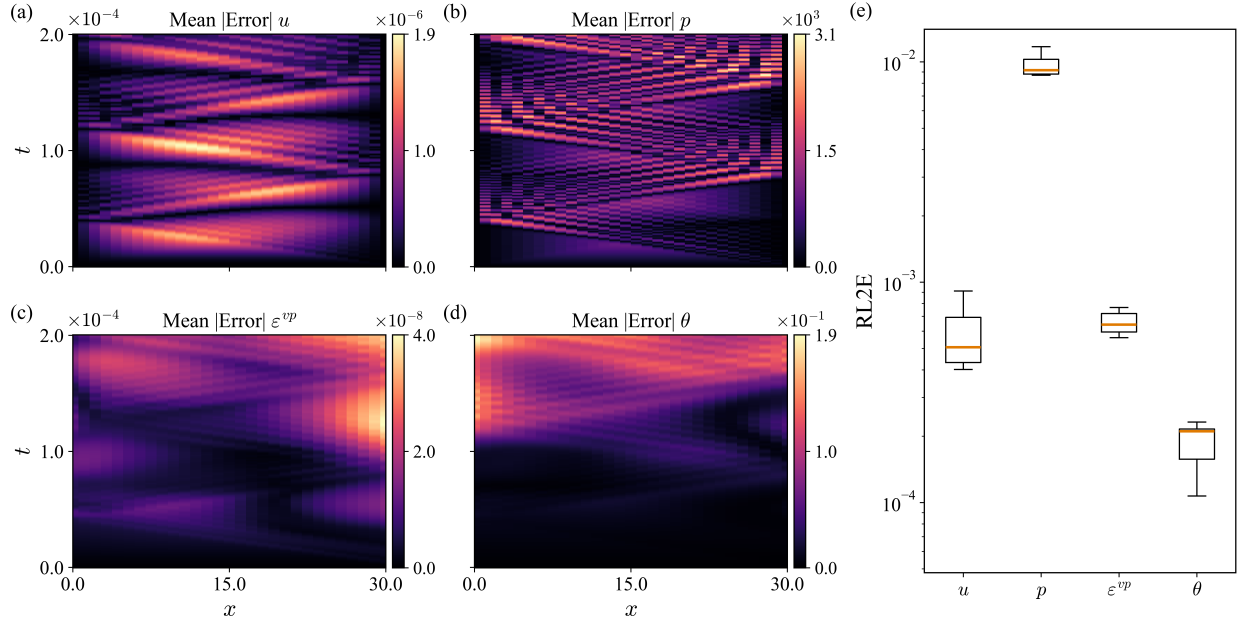}
    \caption{Error statistics for 1D Perzyna viscoplasticity across the full test set. (a)--(d) Mean absolute error fields for (a) displacement~$u$, (b) momentum~$p$, (c) viscoplastic strain $\varepsilon^{vp}$, (d) temperature $\theta$, and (e) boxplots of the trajectory-wise relative $\ell^2$ error (RL2E) for all state variables.}
    \label{fig:example3_error_plot}
\end{figure}

We therefore focus on the predictive performance of the GENERIC model on this system, which is assessed in Figs.~\ref{fig:example3_field_plot} and~\ref{fig:example3_error_plot}. Fig.~\ref{fig:example3_field_plot} presents the spatiotemporal fields for a representative test trajectory. The learned model recovers the qualitative structure of all four fields, with trajectory-wise relative $\ell^2$ errors of $7.24\times10^{-4}$ for $u$, $1.03\times10^{-2}$ for $p$, $1.28\times10^{-3}$ for $\varepsilon^{vp}$, and $3.14\times10^{-4}$ for $\theta$. The largest discrepancies appear in $p$ near the sharp stress-wave fronts, where spatial gradients are steepest. The pointwise error maps (bottom row) confirm that these discrepancies remain spatially localized rather than spreading over time. Fig.~\ref{fig:example3_error_plot} continues this assessment to the full test set. The mean absolute error fields in panels~(a)--(d) reveal that errors concentrate near wave-front regions and do not grow unboundedly in time. The boxplots in panel~(e) show that the median relative $\ell^2$ error is highest for $p$ at order~$10^{-2}$, followed by $u$, $\varepsilon^{vp}$ and $\theta$ at order~$10^{-3}$. The tight interquartile ranges across all variables indicate that the model generalizes well over the sampled boundary conditions. This example validates that the GENERIC framework extends from ODE systems to coupled, nonlinear thermomechanical PDEs successfully. In~\ref{sec:nn_train_details}, we include more details about the training, inference, and model architectures used in this example.

\section{Conclusion}
This paper addresses the problem of thermodynamically-consistent model discovery from data by learning its nonlinear GENERIC structure, i,e., allowing for a general, potentially non-quadratic dissipation potential. The GENERIC building blocks (the reversible operator, the energy, the entropy, and the dissipation potential) are mappings encoded by neural networks, whose architectures are designed to automatically ensure the satisfaction of the first and second laws of thermodynamics (energy conservation and non-negative entropy production).

The proposed framework, termed Nonlinear GENERIC-Embedded Neural Networks (N-GENNs), is illustrated on three examples of increasing complexity: a damped harmonic oscillator, an idealized chemical motor, and a one-dimensional viscoplasticity model. In all three examples, the learned GENERIC models accurately capture the dynamics of the system, while strictly satisfying the first and second laws of thermodynamics. The damped harmonic oscillator serves as a proof of concept, the chemical motor example generalizes it to systems with a noncanonical reversible operator (due to the presence of entropy) and non-quadratic dissipation potential (arising from chemical reactions), and the viscoplasticity example further extends the method to spatiotemporal PDE dynamics. Together, these results demonstrate the flexibility of the proposed approach for learning thermodynamically consistent models across a range of physical settings. These numerical examples should be understood as validation benchmarks designed to assess the proposed methodology, rather than as demonstrations on full-scale experimental multiphysics datasets.

Several limitations remain in the present N-GENNs formulation. First, the method is developed as a deterministic model-recognition framework trained from prescribed state trajectories, and therefore returns a single rollout prediction after training. Second, the numerical examples considered here assume that the relevant thermodynamic state variables are known and observed, whereas many applications involve internal variables, partial observations, or model-form ambiguity. Third, the scalability of the architecture to high-dimensional systems, limited data coverage, and strongly noisy or stochastic trajectories remains to be investigated. These limitations point to the need for uncertainty quantification (UQ), both to assess confidence in the learned model and to distinguish epistemic uncertainty due to sparse or noisy observations, and to incorporate automatic state variable identification strategies. Related uncertainty-aware approaches to thermodynamic learning have recently been explored in \cite{he2026evodms, he2025spiediff}, while the integration of thermodynamic model discovery with automatic strategies for identifying internal variables have been explored in \cite{liu2023learning,rosenkranz2024viscoelasticty,qiu2025bridging,jones2026hierarchy}. In terms of scalability, data-driven reduced order modeling approaches \cite{hernandez2021deep,park2024tlasdi, he2026thermodynamically} offer a promising route by combining autoencoder-based nonlinear dimension reduction with GENERIC-structured latent dynamics.

An important next step will therefore be to extend the design of the current framework to handle high-dimensional multi-physics systems with noisy, sparse, and partially observed measurements. Future work will also focus on extending the framework to incorporate additional structural constraints directly into the architecture, such as the Jacobi identity for the Poisson structure and the Onsager–Casimir reciprocal relations \cite{onsager1931reciprocal, casimir1945onsager}, or their generalizations \cite{pavelka2018multiscale}. Another promising direction is the integration with statistical-mechanical principles to address the inherent non-uniqueness of the inverse problem in model discovery \cite{huang2025statistical}.

\appendix
\section{Neural-network settings, training, and inference details}
\label{sec:nn_train_details}

Table~\ref{tab:nn_architectures} lists the neural-network architectures and optimization settings used to learn the GENERIC components in all three examples. The learned components are reported by their model symbols, $E_{\theta}$, $S_{\psi}$, $\tilde{L}_{\phi}$, and $\tilde{\Xi}_{\omega}$. Each network is an MLP with two hidden layers of $30$ neurons each, using the SoftPlus activation function. The only structural exception is Example 3 (Section~\ref{sec:viscoplasticity}), where the Poisson operator is fixed to its canonical form, and therefore, $\tilde{L}_{\phi}$ is not trained (denoted by ``-'' in Table~\ref{tab:nn_architectures}). We also recall that in Example 3, it is the densities that are learned, as opposed to the functional, and, for the energy, the neural network only approximates the non-kinetic contribution.

For optimization, all experiments use full-batch training (i.e., each update is computed over the entire training dataset) and employ the SOAP optimizer \cite{vyas2025soap} with a learning rate of $3\times 10^{-3}$. In our experiments, SOAP improves optimization stability and convergence behavior, consistent with recent observations in physics-informed machine learning \cite{wang2025gradient}. Since the official code for SOAP is built for PyTorch, here, we implement the unofficial version of SOAP in JAX based on the paper and the official PyTorch code. Interested readers can find this code in the GitHub repository for this project.

Finally, at test time, model rollouts for all examples are integrated using the fourth-order Runge--Kutta (RK4) scheme with the same time step as the corresponding generated dataset: $\Delta t=0.015$ for Example~1 (Section~\ref{sec:harmonic_oscillator}), $\Delta t=0.01$ for Example~2 (Section~\ref{sec:chemical_motor}),  and $\Delta t\approx6.65\times10^{-7}\,\mathrm{s}$ for Example~3. For the last example (Section~\ref{sec:viscoplasticity}), the learned model is rolled out on the same one-dimensional finite-element mesh used for generating the data set, namely a uniform mesh with $\Delta x\approx6.90\times10^{-3}\,\mathrm{m}$. The displacement $u$ and momentum $p$ are represented as nodal degrees of freedom, while $\varepsilon^{vp}$ and $\theta$ are represented as cell-centered element quantities. The rollout states consist of the interior nodal values of $(u,p)$ together with all cell-centered values of $(\varepsilon^{vp},\theta)$; the boundary values of $u$ and $p$ are imposed from the reference boundary histories. In all examples, the representative test trajectory is chosen as the first trajectory in the corresponding test dataset after the random train-test split using the fixed seed for reproducibility. 

\begin{table}[htbp]
    \centering
    \small
    \caption{Neural-network architectures and training hyperparameters used to learn the GENERIC components in the three examples. For each example, we report the hidden-layer width and the number of hidden layers, together with the learning rate and the number of training epochs.}
    \label{tab:nn_architectures}
    \begin{tabular}{c cc cc cc}
        \toprule
        GENERIC & \multicolumn{2}{c}{Example 1} & \multicolumn{2}{c}{Example 2} & \multicolumn{2}{c}{Example 3} \\
        \cmidrule(lr){2-3} \cmidrule(lr){4-5} \cmidrule(lr){6-7}
        Component & Width & Layers & Width & Layers & Width & Layers \\
        \midrule
        $E$ & $30$ & $2$ & $30$ & $2$ & $30$ & $2$ \\
        $S$ & $30$ & $2$ & $30$ & $2$ & $30$ & $2$ \\
        $\tilde{L}$ & $30$ & $2$ & $30$ & $2$ & - & - \\
        $\tilde{\Xi}$ & $30$ & $2$ & $30$ & $2$ & $30$ & $2$ \\
        \midrule
        Learning rate & \multicolumn{2}{c}{$3\times 10^{-3}$} & \multicolumn{2}{c}{$3\times 10^{-3}$} & \multicolumn{2}{c}{$3\times 10^{-3}$} \\
        Num. of epochs & \multicolumn{2}{c}{$10,000$} & \multicolumn{2}{c}{$10,000$} & \multicolumn{2}{c}{$10,000$} \\
        \bottomrule
    \end{tabular}
\end{table}

The neural networks are implemented in Python using JAX and Flax. Standard libraries, including NumPy, SciPy, and Matplotlib, are used for data processing and visualization. Training and inference of the neural networks are performed on a single NVIDIA RTX A6000 GPU.

\section{Example 2 derivation}
\label{sec:motor_derivation}
Starting from the entropy above and setting $R = h = N_A = 1$, the working dimensionless entropy derivatives become $n^*_\mu = \frac{\partial S}{\partial n_\mu}$. Employing the notation $\sum_\beta n_\beta = n$ and matrix multiplication with vector $\bb$ and matrix~$\aaa$, we obtain
\begin{equation}
        n^*_\mu = \ln\left(\frac{Aq - \nn^T \bb}{n_\mu }\left(\frac{4\pi m_\mu (\epsilon + \frac{\nn^T \aaa \nn}{Aq})}{3 n}\right)^{\frac{3}{2}}\right) - \frac{b_\mu n}{Aq - \nn^T \bb} + \frac{3n\sum_{\sigma} a_{\mu\sigma} n_\sigma}{\epsilon Aq + \nn^T \aaa \nn}
\end{equation}
where we have used the symmetry of the interaction $\aaa = \aaa^T$. This gives us the thermodynamic force
\begin{multline}
    n^*_1 + n^*_2 - n^*_3 = \frac{3n \sum_{\sigma} (a_{1\sigma} + a_{2\sigma} - a_{3\sigma})n_\sigma }{\epsilon Aq + \nn^T \aaa \nn} - \frac{n(b_1 + b_2 - b_3)}{Aq - \nn^T \bb } \\ + \ln\left(\frac{n_3}{n_1 n_2}\bigg[\frac{m_1 m_2}{m_3}\bigg]^{\frac{3}{2}} (Aq - \nn^T \bb)\left(\frac{4\pi (\epsilon + \frac{\nn^T \aaa \nn}{Aq})}{3 n}\right)^{\frac{3}{2}}\right).
\end{multline}
And through derivatives of the dissipation potential, the reaction rates
\begin{align}
    \dot{n}_1 &= \frac{\alpha}{4}(Kn_3 - \frac{1}{K}n_1 n_2) ,\nonumber\\
    \dot{n}_2 &= \frac{\alpha}{4}(Kn_3 - \frac{1}{K}n_1 n_2) ,\\
    \dot{n}_3 &= \frac{\alpha}{4}(\frac{1}{K}n_1 n_2 - Kn_3). \nonumber
\end{align}
where we denote by $K$ the expression
\begin{equation}
    K = \exp\bigg[\frac{3n \sum_{\sigma} (a_{1\sigma} + a_{2\sigma} - a_{3\sigma})n_\sigma }{2(\epsilon Aq + \nn^T \aaa \nn)} - \frac{n(b_1 + b_2 - b_3)}{2(Aq - \nn^T \bb)}\bigg]\left(\frac{m_1 m_2}{m_3}\right)^{\frac{3}{4}} (Aq - \nn^T\bb)^{\frac{1}{2}}\left(\frac{4\pi (\epsilon + \frac{\nn^T\aaa\nn}{Aq})}{3 n}\right)^{\frac{3}{4}}.
\end{equation}

Relating the force $F$ to the area of the piston $A$, we can write $F(\xx) = P(\xx)A$. We can then calculate the pressure from the derivatives of entropy. This gives us the well known van der Waals equations of state
\begin{align}
    P + \frac{\sum_{\rho\sigma}n_\rho n_\sigma a_{\rho\sigma}}{V^2} = \frac{RT\sum_\beta n_\beta}{V - \sum_\nu n_\nu b_\nu}\label{eq:thermal}, \\
    \epsilon = \frac{3}{2}RT\sum_\beta n_\beta - \frac{\sum_{\rho\sigma}n_\rho n_\sigma a_{\rho\sigma}}{V} \label{eq:calor}
\end{align}
Substituting for temperature $T$ from \eqref{eq:calor} to \eqref{eq:thermal} we obtain the final form of the equations of motion as
\begin{align}
    \dot{q} &= \frac{p}{m}, \nonumber\\
    \dot{p} &= -k(q-q_0) + \frac{2}{3}\frac{\epsilon Aq + \sum_{\rho\sigma}n_\rho n_\sigma a_{\rho\sigma}}{q(Aq - \sum_\nu n_\nu b_\nu)} - \frac{\sum_{\rho\sigma}n_\rho n_\sigma a_{\rho\sigma}}{Aq^2},\nonumber\\
    \dot{\epsilon} &= -\frac{p}{m}\left(\frac{2}{3}\frac{\epsilon Aq + \sum_{\rho\sigma}n_\rho n_\sigma a_{\rho\sigma}}{q(Aq - \sum_\nu n_\nu b_\nu)} - \frac{\sum_{\rho\sigma}n_\rho n_\sigma a_{\rho\sigma}}{Aq^2}\right), \nonumber\\
    \dot{n}_1 &= \frac{\alpha}{4}(Kn_3 - \frac{1}{K}n_1 n_2) ,\\
    \dot{n}_2 &= \frac{\alpha}{4}(Kn_3 - \frac{1}{K}n_1 n_2) ,\nonumber\\
    \dot{n}_3 &= \frac{\alpha}{4}(\frac{1}{K}n_1 n_2 - Kn_3). \nonumber
\end{align}
\subsection{Second degeneracy condition and the Jacobi identity}
Lastly, the degeneracy condition $L \DD S = 0$ and the Jacobi identity must be satisfied. Since $S(\xx)$ does not depend on $p$, the only equation we need to check for degeneracy is whether it holds that
\begin{equation}
    \frac{\partial S}{\partial \epsilon}F(\xx) = \frac{\partial S}{\partial q}
\end{equation}
However, this is obvious through our construction of the force $F$, since from the fundamental thermodynamic relation $\dd S = \frac{\dd\epsilon}{T} + \frac{P}{T}\dd V - \sum_{\mu=1}^3 \frac{\mu_\mu}{T}\dd n_\mu$ we have
\begin{equation}
    \frac{1}{T}F(x) = A\frac{P}{T}
\end{equation}
which is automatically satisfied for $F = PA$.

The Jacobi identity is
\begin{equation}
    \sum^6_{m=1}L_{im}\frac{\partial L_{jk}}{\partial x_m} + L_{jm}\frac{\partial L_{ki}}{\partial x_m} + L_{km}\frac{\partial L_{ij}}{\partial x_m} = 0.
\end{equation}
If any of the indices $i,j,k$ were greater than $3$, the corresponding components of $L$ would vanish; see structure of $L$ in \eqref{eq:motor_energy}. Additionally, if any of the two free indices were to repeat, we would get zero due to the skew-symmetry of $L$.  Finally, if the identity holds for a given triple $(i,j,k)$, it also holds for any other permutation of these indices. Consequently, it suffices to verify the relation for the single nontrivial case $(i,j,k) = (1,2,3)$. 
\begin{equation}
        \sum^6_{m=1}L_{1m}\frac{\partial L_{23}}{\partial x_m} + L_{2m}\frac{\partial L_{31}}{\partial x_m} + L_{3m}\frac{\partial L_{12}}{\partial x_m} =  \sum^6_{m=1}L_{1m}\frac{\partial L_{23}}{\partial x_m} = \frac{\partial F(q, \epsilon, \nn)}{\partial p} = 0.
\end{equation}
Therefore, the Jacobi identity is satisfied.

\section{Discretization scheme in Example 3} \label{appx:example3_discretization}
The governing equations are discretized using a finite element scheme on a uniform mesh with spacing $h$ and a fixed time step $\Delta t$. For each field, time levels are denoted by superscripts, while spatial nodes are indexed by subscripts, e.g., $u^n_i$ denoting the displacement at position $x_i$ and time $t^n$. Linear shape functions $N_i$ are employed, satisfying the standard delta-Kronecker property $N_i(x_j)=\delta_{ij}$.

The displacement $u$ and momentum $p$ are defined at the nodal points, whereas the viscoplastic strain $\varepsilon^{vp}$ and temperature $\theta$ are evaluated at Gauss points, located at the center of each element. Under this discretization, the strain $\varepsilon$ and associated stress $\sigma$ are constant within each element and can therefore also be treated as a Gauss-point quantities.

The weak form of the evolution equation for $u$ (resulting from combining the first two equations in \eqref{eq: example 3}) is given by
\begin{equation}
    \int_0^L N_i \rho \ddot{u} \,\dd x 
    = -\int_0^L \frac{\partial N_i}{\partial x} C \left( \frac{\partial u}{\partial x} 
    -  \varepsilon^{vp} \right) \,\dd x,
\end{equation}
where $i$ denotes any index associated to an interior node. Upon integration, this leads to 
\begin{equation}
    \frac{\rho}{6} \left( \ddot{u}^n_{i+1} + 4\ddot{u}^n_{i} + \ddot{u}^n_{i-1} \right) 
    = C \frac{u^{n}_{i+1}-2u^{n}_{i}+u^{n}_{i-1}}{h^2} 
    - C \frac{\varepsilon^{vp,n}_{i+1/2} - \varepsilon^{vp,n}_{i-1/2}}{h}.
\end{equation}
For simplicity, a lumped mass approximation is adopted, yielding
\begin{equation}
    \rho \ddot{u}^n_{i} 
    = C \frac{u^{n}_{i+1}-2u^{n}_{i}+u^{n}_{i-1}}{h^2} 
    - C \frac{\varepsilon^{vp,n}_{i+1/2} - \varepsilon^{vp,n}_{i-1/2}}{h}.
\end{equation}
The nodal displacement values are updated using a velocity-Verlet algorithm, i.e.
\begin{align}
&u_i^{n+1}
= u_i^n + \Delta t\, v_i^n + \frac{\Delta t^2}{2} a_i^n, \\
&v_i^{n+1}
= v_i^n + \frac{\Delta t}{2}\left(a_i^n + a_i^{n+1}\right),
\end{align}
where $v_i^n = \dot{u}^n_{i}$ is the velocity, and $a_i^n = \ddot{u}^n_{i}$ is the acceleration. The viscoplastic strain $\varepsilon^{vp}$ and temperature $\theta$ follow first-order differential equations (see Eq.~\eqref{eq: example 3}), which are discretized in time using a forward Euler scheme
\begin{equation}
    \frac{\varepsilon^{vp,n+1}_{i+1/2} - \varepsilon^{vp,n}_{i+1/2}}{\Delta t} = \frac{1}{\eta} \left\langle \Big| \sigma_{i+1/2}^n \Big| - \sigma_y \right\rangle \text{sign} \left( \sigma_{i+1/2}^n \right)
\end{equation}
\begin{equation}
    \frac{\theta_{i + 1/2}^{n+1} - \theta_{i + 1/2}^{n}}{\Delta t} =\sigma_{i+1/2}^n \frac{1}{c} \frac{\varepsilon^{vp,n+1}_{i+1/2} - \varepsilon^{vp,n}_{i+1/2}}{\Delta t},
\end{equation}
with
\begin{equation}
 \sigma_{i+1/2}^n=C \left(\frac{u^n_{i+1} - u^n_i}{h} - \varepsilon^{vp,n}_{i+1/2} \right).
\end{equation}

\section*{Code availability}
The code used to generate the results in this paper is publicly available on GitHub at \url{https://github.com/celiareina/N-GENNs}.

\section*{Declaration of competing interest}
The authors declare that they have no known competing financial interests or personal relationships that could have appeared to influence the work reported in this paper.

\section*{Acknowledgments}
M.P. and V.V.~were supported by the Czech Science Foundation, project 23-05736S. M.P. is a member of the Nečas Center of Mathematical Modeling. Z.H., W.Q.,~and C.R.~gratefully acknowledge support from NSF CAREER Award, CMMI-2047506.

\section*{Declaration of generative AI and AI-assisted technologies in the writing process}
During the preparation of this work, the authors used ChatGPT in order to check the grammar and improve the clarity of sentences. After using this tool/service, the authors reviewed and edited the content as needed and take full responsibility for the content of the published article.

\bibliographystyle{abbrv}
\bibliography{references.bib}

\end{document}